%% file: Paper.tex
\documentclass[3p,10pt,authoryear]{elsarticle}

\usepackage{adjustbox}
\usepackage{graphicx,algorithm,algorithmicx,amsmath}
\usepackage{amssymb}
\usepackage{graphics}
\usepackage{epsfig}
\usepackage{epstopdf}
\usepackage{multirow}
\usepackage{multicol}

\renewcommand{\dbltopfraction}{0.95}	
\usepackage{array}
\newcommand{\stackcell}[2][c]{%
 \begin{tabular}{@{}#1@{}}
 #2
 \end{tabular}%
}
\usepackage{stfloats}
\usepackage{float}
\usepackage[usenames,dvipsnames]{color}

\usepackage{helvet}
\usepackage{tabularx}

\usepackage{balance,setspace}

\begin{document}

\begin{frontmatter}



\title{A Bayesian Approach to Estimation of Speaker Normalization Parameters}

\author[idiap,epfl]{Dhananjay Ram\corref{cor1}}
\cortext[cor1]{Corresponding author}
\author[iitk]{Debasis Kundu} 
\author[iitk]{Rajesh M. Hegde}

\address[idiap]{Idiap Research Institute, Martigny, Switzerland}
\address[epfl]{Ecole Polytechnique F\'ed\'erale de Lausanne (EPFL), Lausanne,
Switzerland}
\address[iitk]{Indian Institute of Technology Kanpur, India}

\address{dhananjay.ram@idiap.ch}
\address{\{kundu,rhegde\}@iitk.ac.in}

\begin{abstract}
In this work, a Bayesian approach to speaker normalization is proposed to compensate for the degradation in performance of a speaker independent speech recognition system. The speaker normalization method proposed herein uses the technique of vocal tract length normalization (VTLN). The VTLN parameters are estimated using a novel Bayesian approach which utilizes the Gibbs sampler, a special type of Markov Chain Monte Carlo method. Additionally the hyperparameters are estimated using maximum likelihood approach. This model is used assuming that human vocal tract can be modeled as a tube of uniform cross section. It captures the variation in length of the vocal tract of different speakers more effectively, than the linear model used in literature. The work has also investigated different methods like minimization of Mean Square Error (MSE) and Mean Absolute Error (MAE) for the estimation of VTLN parameters. Both single pass and two pass approaches are then used to build a VTLN based speech recognizer. Experimental results on recognition of vowels and Hindi phrases from a medium vocabulary indicate that the Bayesian method improves the performance by a considerable margin.
\end{abstract}


\end{frontmatter}

\input{introduction}
\input{background}
\input{Bayesian_Approach}
\input{experiments}
\input{conclusion}
\input{appendix}





\bibliographystyle{elsarticle-num}
\balance
\bibliography{thesis}

%
%
%

\end{document}

%% file: introduction.tex
\section{Introduction}
\label{sec:intro}
One of the biggest challenges in the design of an automatic speech recognizer (ASR) is to compensate for the speaker variability. It is caused by the acoustic variations introduced in the signals of same utterance spoken by different speakers. In spite of these acoustic variations, humans can recognize utterances of different speakers very easily. But, speech recognizer cannot recognize same words uttered by different speakers very well due to these variations \cite{o2003interacting}. Different types of speaker normalization methods are used to enhance the recognition accuracy when different speakers are using same recognizer \cite{cohen1994experiment, lee1996speaker, umesh2002simple}. Many researchers have approached the problem of normalization using only formants of vowels. Nordstrom and Lindblom \cite{nordstrom1975} determined a constant scale factor depending on the ratio of third formant of subject to that of reference speaker. Later, Fant \cite{fant1975non} argued that uniform scaling is a very simple approximation, so the scale factor should be dependent on both vowel category and formant number. Then, Miller \cite{miller1989auditory} applied formant ratio theory to normalization problem, which claims that vowels are relative patterns. A detailed study of these vowel normalization procedures by Nicholas Flynn can be found in \cite{flynn2011comparing}.

Warping functions are also used to reduce differences between the spectra of subject and reference speakers. Different types of warping functions have been used in past, namely Linear warping function \cite{lee1998frequency}, Piecewise Linear warping function \cite{wegmann1996speaker}, Affine warping function \cite{sinha2002non}, Non-linear warping function \cite{umesh1999fitting} etc. Eide and Gish \cite{eide1996parametric} have developed a warping function based on the median position of third formant in speech of the speaker under consideration. Stevens and Volkmann \cite{melscale1940} have proposed a non-linear warping function based on their perceptual studies on speech signals. The well known bilinear transform is also used as a warping function in \cite{acero1991robust}. A likelihood maximization technique of speaker normalization is proposed in \cite{cohen1994experiment}. In this method, warping factor for a particular speaker is chosen by maximizing the likelihood of a hypothesis in an iterative manner at the output of a recognizer. In \cite{wegmann1996speaker}, Gaussian Mixture Model is used to represent a class of standard speakers. The classes are assigned warping factors from a set of values. Different speaker are assigned to one of these classes using likelihood maximization. Lee and Rose \cite{lee1998frequency} have proposed a similar method. But, they have used maximization of likelihood of the Hidden Markov Model (HMM) for normalization as well as recognition. Researchers have also used normalization in the feature domain. Acero and Stern \cite{acero1991robust} have proposed an affine transformation of the cepstral features. Cox \cite{cox2000speaker} has shown that vocal tract length normalization (VTLN) can be implemented with filter banks and directly applied this method in the feature domain. Later in \cite{sanand2007linear}, a linear transformation approach using Dynamic Frequency Warping (DFW) has been proposed for normalization.

In this paper an affine model based speaker normalization method is presented. The model parameters are estimated using Bayesian estimation as well as error function minimization technique. This work justifies the use of Bayesian method of parameter estimation, which is also supported by the experimental results. Techniques like bandwidth adjustment and frequency bin adjustment are discussed which are required for the application of speaker normalization in the speech recognizer. The rest of the paper is organized as follows. Section \ref{sec:speaker_norm} describes the model and error function minimization technique is used to estimate the model parameters. Section \ref{sec:bayesian} proposes Bayesian estimation method to estimate the affine model parameters. The need of hyperparameters estimation is also discussed in this Section and the hyperparameters are estimated using likelihood maximization. Section \ref{sec:PerEv} presents Formant frequency based vowel recognizer and HMM based speech recognizer, and experimental condition to perform different types of experiments on these recognizers. The incorporation of normalization method with recognizer for training as well as testing on a database is described using block diagrams. 
Experimental results for both vowel as well as word recognition are shown in the same Section. Finally, conclusions are presented in Section \ref{sec:Con}.


%% file: background.tex
\section{Speaker Normalization using Frequency Warping}
\label{sec:speaker_norm}

An affine model \cite{kumar2008nonuniform} for speaker normalization is introduced in this section. The model parameters, $\alpha$ and $\kappa$ are estimated using principle of Error Function Minimization. However it is found that, for a range of values of $\alpha$, estimated value of $\kappa$ is not reliable. An adjustment technique is suggested for estimated values of parameters to get more reliable estimates.


\subsection{Affine Model for Vocal Tract Length Normalization}
\label{sec:model}
The affine model used for speaker normalization in an earlier work \cite{kumar2008nonuniform} is given by
\begin{equation}
{\bf Y}=\alpha{\bf X}+\kappa(\alpha-1){\bf 1}
\label{eq:model}
\end{equation}
where, the vectors, ${\bf Y}$ and ${\bf X}$ represent the formant frequency vector for reference and subject speakers respectively. Formant frequency vector is constructed by concatenating formants of all utterances corresponding to a particular speaker. So, length of this vector depends on the database under consideration.
$\alpha$ and $\kappa$ are respectively, the speaker dependent and independent parameters.
${\bf 1}$ is a vector of $1$s, i.e. ${\bf 1}$ = $[1\:1\:.....\:1]^T$. The dimension of ${\bf 1}$ is same as the dimension of ${\bf Y}$ or ${\bf X}$.

\begin{figure}[htb]
\begin{minipage}[b]{0.48\linewidth}
  \centering
  \centerline{\includegraphics[scale=0.4]{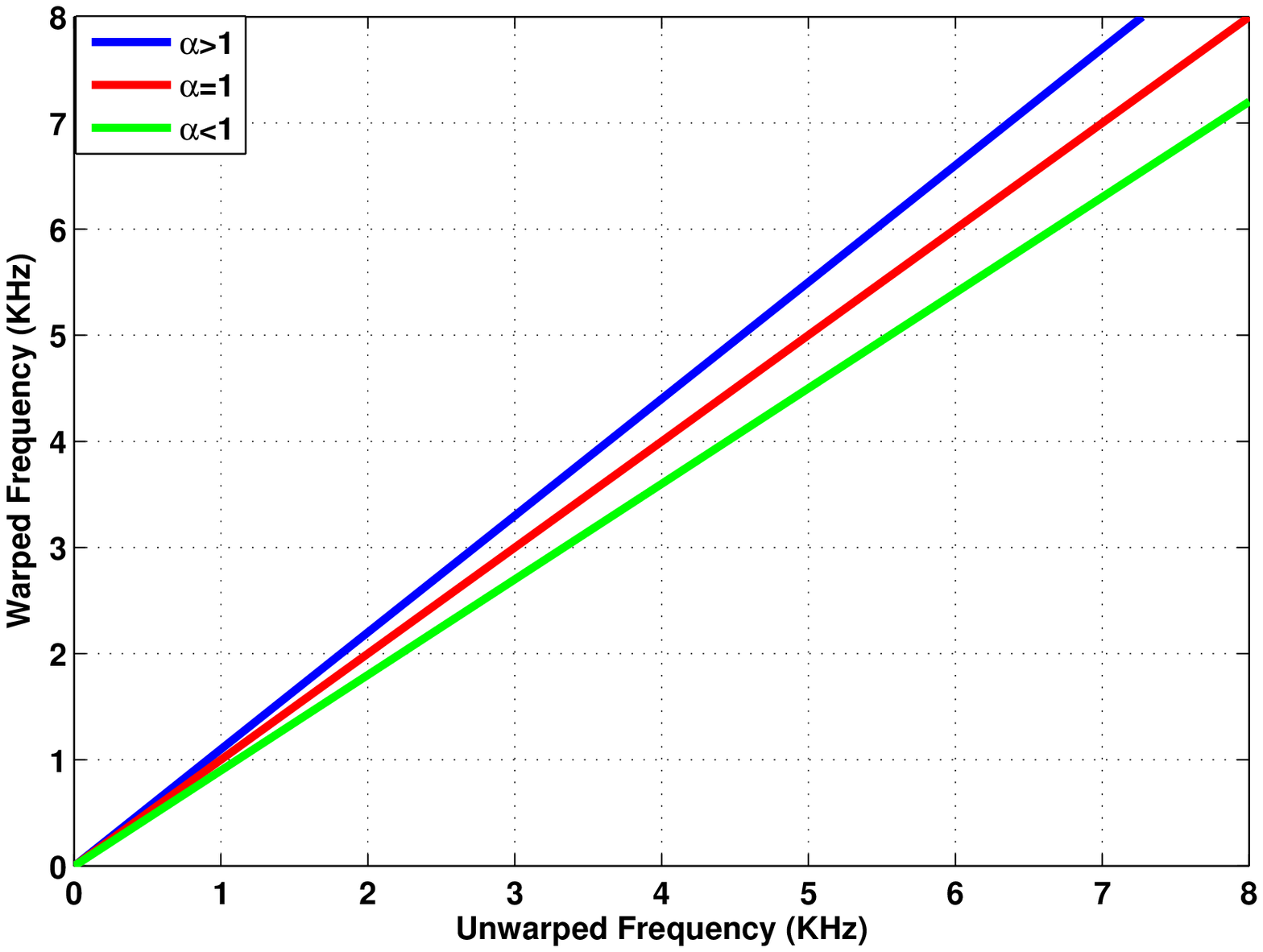}}
  \centerline{(a) Linear Model}
\end{minipage}
\begin{minipage}[b]{0.48\linewidth}
  \centering
  \centerline{\includegraphics[scale=0.4]{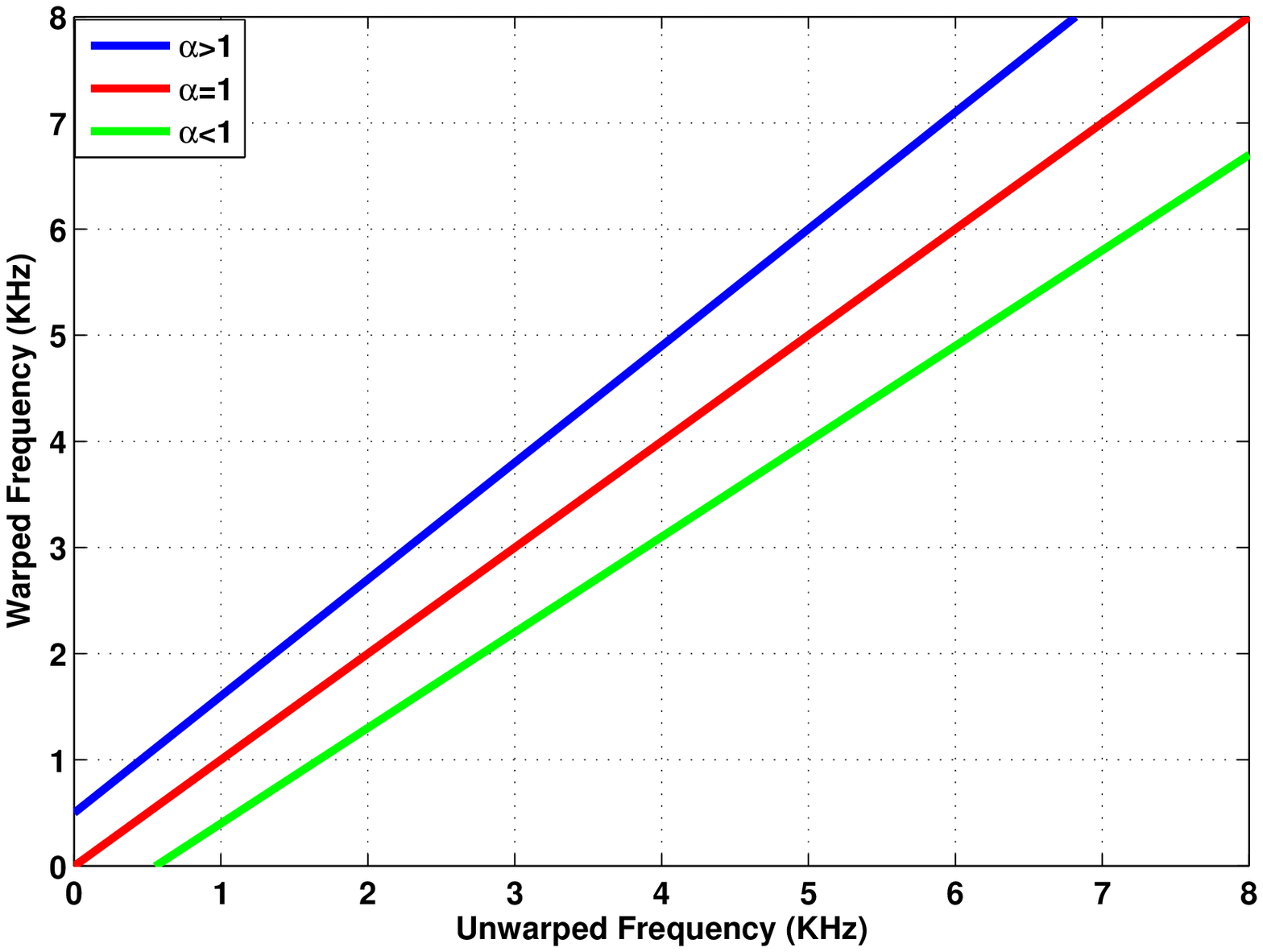}}
  \centerline{(b) Affine Model}
\end{minipage}
\caption{Comparison of warping functions for different values of $\alpha$ and constant $\kappa$}
\label{fig:model}
\end{figure}

It can be easily seen from Figure \ref{fig:model}, that the affine model has a shift factor in addition to the scaling factor used in the popular linear model for VTLN. The linear warping function always passes through the origin, whereas the affine warp function passes through the origin only for  $\alpha=1$. In other cases, it has a positive intercept for $\alpha>1$ and negative intercept for $\alpha<1$. The warped spectrum gets shifted due to non-zero intercept of the normalization function. The shift is towards right or left for values of $\alpha$ greater than 1 or less than 1 respectively.

The shift factor in Equation (\ref{eq:model}) is not same for different subject speakers. But, the number of parameters has not been doubled to achieve this, compared to the number of parameters in the linear model. It is achieved simply by making the shift factor, a function of the speaker dependent scaling factor. In this way, the increase in the number of parameters is only one, but an effect of as many parameters as the number of speakers has been achieved. Thus, a lesser accurate model is obtained compared to the scenario when the shift factors are also speaker dependent . It is implemented to keep a check on the number of parameters to be estimated, which effectively reduces computational complexity of the system. This same affine model was used in \cite{kumar2008nonuniform}. But, the authors have used it to come up with a universal warping function having the same parametric form as the mel scale.

\begin{figure}[!h]
 \centering
 \centerline{\includegraphics[height=7.5cm, width=13cm]{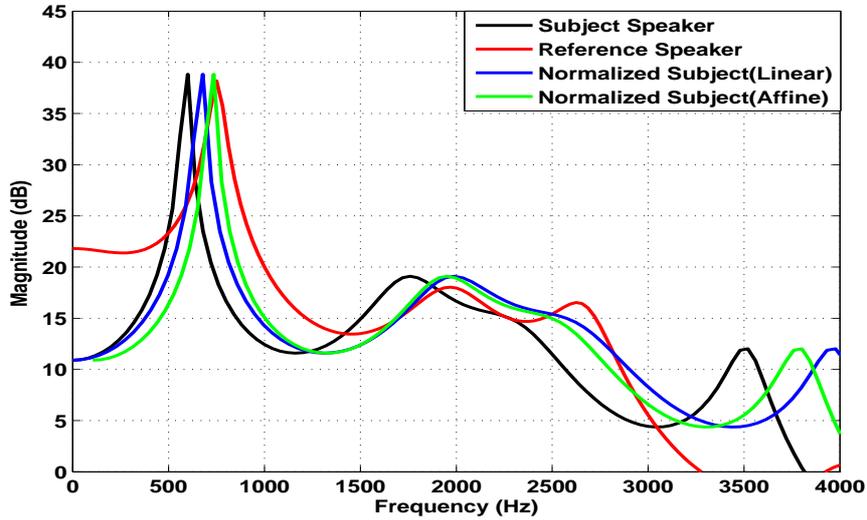}}
 \vspace{-2mm}
 \caption{Normalization of spectrum of the vowel `ae'}
\label{fig:normlm}
\end{figure}

Figure \ref{fig:normlm} shows four different LPC smoothed spectra of same vowel, `ae'. In this figure, two spectra represent utterances by a subject speaker and a reference speaker, whereas the other two spectra are normalized spectra of the subject speaker to make it closer to the reference speaker in terms of formant peaks. The two normalized spectra are obtained using popularly used linear model and an affine model respectively. The first and second formant frequency values are given in Table \ref{table:formant}. In this table, values in first row shows formants of reference speaker. The following rows indicate formant of normalized spectra of subject speaker using linear and affine model, respectively.

\begin{table}[!h]
\begin{center}
 \caption{Table Showing First Two Formant Frequencies of Spectra Presented in Figure \ref{fig:normlm}}
 \label{table:formant}
 \begin{tabular}{|c|c|c|}
 \hline
 \multirow{2}{2.5cm}{\bfseries\stackcell{Normalization\\ Function}} & \multicolumn{2}{c|}{\bfseries Formant Frequency}
 \tabularnewline
 \cline{2-3}
 & {\bfseries First Formant (Hz)} & {\bfseries Second Formant (Hz)}
 \tabularnewline
 \hline
 Reference & 750 & 1970 \tabularnewline
 \hline
 Subject & 600 & 1760 \tabularnewline
 \hline
 Linear Function & 678 & 2000 \tabularnewline
 \hline
 Affine Function & 735 & 1960 \tabularnewline
 \hline
 \end{tabular}
\end{center}
\end{table}

The formant frequency values of normalized spectra clearly show that, affine model gives a better match compared to the linear model, in terms of formant frequencies.

\subsection{Affine Model Parameter Estimation using Error Function Minimization Technique}
\label{sec:efmt}
An error function minimization technique is used for estimating the parameters $\alpha$ and $\kappa$ of the affine model mentioned earlier. The model can be written as
\begin{equation}\label{eq:errmodel}
{\bf Y}=\alpha{\bf X}+\kappa(\alpha-1){\bf 1}+{\boldsymbol\epsilon}
\end{equation}
where, ${\boldsymbol \epsilon}$ is the error vector. The estimated formant frequency vector after applying these parameters on ${\bf X}$ is
\begin{equation}\label{eq:estimate}
 \widehat{{\bf Y}}=\alpha{\bf X}+\kappa(\alpha-1){\bf 1}
\end{equation}
The parameter $\kappa$ is assumed to be speaker dependent. Then, the estimated formant frequency vector can be written as
\begin{equation}\label{eq:est_y_ij}
 \widehat{{\bf Y}}_{ij}=\alpha_{ij}{\bf X}_{j}+\kappa_{ij}(\alpha_{ij}-1){\bf 1}
\end{equation}
where $\alpha_{ij}$ and $\kappa_{ij}$ are the parameters for the $j$-th subject speaker with respect to $i$-th reference speaker.

The above Equation (\ref{eq:est_y_ij}), when averaged over all the reference speakers gives the following 

\begin{equation}\label{eq:est_y_j}
 \widehat{{\bf Y}}_{j}=\alpha_{j}{\bf X}_{j}+\kappa_{j}(\alpha_{j}-1){\bf 1}
\end{equation}

Now, the parameters, $\kappa_{ij}$ and $\alpha_{ij}$ in Equation (\ref{eq:est_y_ij}) are estimated by considering every possible pair of subject and reference speaker and minimizing an error function. Mean Square Error (MSE) and Mean Absolute Error (MAE) have been used as error functions. \\
In MSE, the error function to be minimized is
\begin{equation}
\begin{aligned}
 {\boldsymbol \epsilon}_{ij} & = (\:{\bf Y}_{ij} - \widehat{{\bf Y}}_{ij})^T(\:{\bf Y}_{ij} - \widehat{{\bf Y}}_{ij}) \\
  & = (\:{\bf Y}_{ij}-\alpha_{ij}{\bf X}_{j}+\kappa_{ij}(\alpha_{ij}-1){\bf 1}\:)^T(\:{\bf Y}_{ij}-\alpha_{ij}{\bf X}_{j}+\kappa_{ij}(\alpha_{ij}-1){\bf 1}\:)
\end{aligned}
 \label{eq:mse}
\end{equation}
In MAE, the error function to be minimized is
\begin{equation}\label{eq:mae}
\begin{aligned}
 {\boldsymbol \epsilon}_{ij} & = |\:{\bf Y}_{ij} - \widehat{{\bf Y}}_{ij}|^T {\bf 1} \\
 & = |\:{\bf Y}_{ij}-\alpha_{ij}{\bf X}_{j}+\kappa_{ij}(\alpha_{ij}-1){\bf 1}\:|^T {\bf 1}
\end{aligned}
\end{equation}
where, the operator $|*|$ gives element-wise absolute value of corresponding vector. Two different estimates of $\kappa_{ij}$ and $\alpha_{ij}$ are obtained by minimizing Equations (\ref{eq:mse}) and (\ref{eq:mae}). For each of the above mentioned functions, joint optimization has been used for estimating $\kappa_{ij}$ and $\alpha_{ij}$. Nelder-Mead method \cite{nelder1965simplex} is implemented to minimize both Equations (\ref{eq:mse}) and (\ref{eq:mae}). This is a well defined numerical method for problem for which derivatives may not be known. Now, it is assumed that, there are $m$ subject speakers and $n$ reference speakers. So, the average estimated formant frequency vector for the $j$-th subject speaker will be
\begin{equation} \label{eq:avg_ref}
 \widehat{{\bf Y}}_{j}=\left(\frac{1}{n}\displaystyle\sum_{i=1}^{n}\alpha_{ij}\right){\bf X}_{j}+ \left(\frac{1}{n}\displaystyle\sum_{i=1}^{n}\kappa_{ij}(\alpha_{ij}-1)\right){\bf 1}
\end{equation}
Now, comparison of Equations (\ref{eq:est_y_j}) and (\ref{eq:avg_ref}) gives the following,

\noindent\begin{minipage}{.45\linewidth}
\begin{equation}
\displaystyle\alpha_{j}=\frac{1}{n}\displaystyle\sum_{i=1}^{n}\alpha_{ij}
\label{eq:alpha_lse}
\end{equation}
\end{minipage}
\begin{minipage}{.45\linewidth}
\begin{equation}
\displaystyle\kappa_{j}=\frac{\displaystyle\sum_{i=1}^{n}\kappa_{ij}(\alpha_{ij}-1)}{\displaystyle\sum_{i=1}^{n}(\alpha_{ij}-1)}
\label{eq:kappa_j}
\end{equation}
\end{minipage}

The shift factor, $\kappa$ for the entire database is obtained by averaging $\kappa_{j}$ over all the subject speakers
\begin{equation} \label{eq:kappa}
 \displaystyle\kappa=\frac{1}{m}\displaystyle\sum_{j=1}^{m}\kappa_{j}
\end{equation}

\begin{figure}
 \centering
 \centerline{\includegraphics[height=6.5cm, width=13cm]{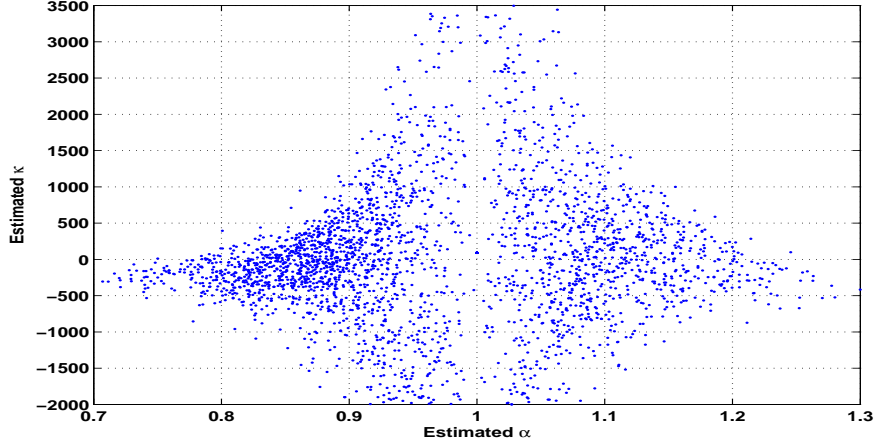}}
 \caption{Illustration of large variability of $\kappa$ near $\alpha=1$}
 \label{fig:avsk}
\end{figure}

An ardent observation of Equation (\ref{eq:model}) indicates that, for $\alpha = 1$, $\kappa$ can take any value from $-\infty$ to $+\infty$. Hence, $\kappa$ cannot be estimated reliably. Also, the estimated values of $\kappa$ show a great amount of variance for values of $\alpha$ close to 1. Figure \ref{fig:avsk} shows the variation of estimated $\kappa$ with corresponding estimates of $\alpha$. In other words, estimated values of $\kappa$ are also not reliable for the values of $\alpha$ near 1. An outlier adjustment techniques is applied to deal with this problem.
An outlier is defined as an estimated value that deviates markedly from other estimated values of a parameter. In this method, a range of $[0,L]$ is chosen for estimated $\kappa_{ij}$, for $i$-th reference speaker and $j$-th subject speaker. The following expression is used to adjust the values of $\kappa_{ij}$ to the boundary.
\begin{equation}
 \kappa_{ij}= max (0, min(\kappa_{ij} , L))
 \label{kappa_adj}
\end{equation}
The adjusted values for $\kappa_{ij}$ are $0$ and $L$ for $\kappa_{ij}<0$ and $\kappa_{ij}>L$ respectively.
The outlier adjustment technique discussed here is an empirical method based on observations. Note that, the MMSE estimation and corresponding outlier adjustment technique has been proposed in \cite{kumar2008nonuniform}, whereas the MMAE estimation technique is proposed in this paper. This technique has been successfully applied in \cite{kumar2008nonuniform} with affine model. The same method is applied here in order to compare its results with the proposed Bayesian method. Also, the results in Section \ref{sec:PerEv} show that, this method improves performance of vowel recognizer for gender dependent normalization in most cases but not for all.

Now, the algorithm to evaluate affine model parameters using error function minimization technique is given below.


\begin{algorithm}
\label{alg1}
\caption{Estimation of Parameters $\alpha$ and $\kappa$ using MMSE and MMAE}
\begin{algorithmic}[1]
\State {\bf {Formant Frequency Vectors}} : Formant frequencies are extracted \cite{snell1993formant} from all utterances to construct fromant frequency vector for subject speaker ({\bf X}) and reference speaker ({\bf Y}).
\State {\bf{Error function}} : MSE and MAE for $i$-th reference speaker and $j$-th subject are constructed as shown in Equations (\ref{eq:mse}) and (\ref{eq:mae}) respectively.
\State {\bf{Estimation of $\alpha_{ij}$ and $\kappa_{ij}$}} : $\alpha_{ij}$ and $\kappa_{ij}$ are estimated by minimizing Equations (\ref{eq:mse}) and (\ref{eq:mae}) suing Nelder-Mead method. These error functions give two different estimate of the same parameter.
\State {\bf{Final $\alpha$}} : Finally, $\alpha_j$ for $j$-th subject speaker is obtained by averaging $\alpha_{ij}$ over all reference speakers.
\State {\bf{Outlier adjustment for $\kappa_{ij}$}} : If $\kappa_{ij}$ values are outside a given range $[0,L]$, its values are adjusted to this range using Equation (\ref{kappa_adj}).
\State {\bf{Final $\kappa$}} : First $\kappa_j$ is calculated using Equation (\ref{eq:kappa_j}) and then, it is averaged over all subject speakers to obtain final $\kappa$ for the database.
\end{algorithmic}
\end{algorithm}
MMSE and MMAE estimation techniques have been used in this section to estimate the affine model parameters. But, these technique require $\kappa$ to be speaker dependent, which is essentially opposite to the premise of the model, where it is assumed to be speaker independent. Also, the huge variance of $\kappa$ is controlled using an outlier adjustment technique, in which the adjustment range may change depending on the database. In order to overcome these problems, Bayesian estimation method is proposed in the following section. 



%% file: Bayesian_Approach.tex
\section{Bayesian Approach to Estimation of Affine Model Parameters}
\label{sec:bayesian}


In this section, Bayesian Estimation has been introduced for estimating the parameters. First, the motivation behind using the Bayesian estimation is discussed. The motivation is followed by an elaborate mathematical description of Bayesian estimation technique for the problem at hand. Additionally, Gibbs sampler is used for numerical estimation of the model parameters. Subsequently, maximum likelihood estimation technique is used to estimate hyperparameters of the model.

\subsection{Motivation}
The proposed Bayesian method for affine model parameter estimation is broadly motivated by the following observations,

\begin{itemize}
\item The speaker independent parameter $\kappa$ of the affine model described in Section \ref{sec:model} is not estimable under the frequentist set-up, if true value of $\alpha=1$. Even if $\alpha \neq 1$ but very close to 1, which is usually the case in practice, the least squares estimate or the maximum likelihood estimate (under the assumption of Gaussian error) of $\alpha$ becomes very unreliable. It has been observed in the simulation study presented in Figure \ref{fig:avsk} that the variance of the estimator of $\kappa$ is very high in such a situation.

\item On the other hand under in Bayesian framework, because of the random nature of $\alpha$, $\kappa$ is always estimable. Since in practice $\alpha$ is very close to 1, if we can use our prior knowledge on $\alpha$, very reliable Bayes estimate of $\kappa$ can be obtained. Even if we do not have any prior knowledge of $\alpha$, the data driven prior works much better than the usual least squares or maximum likelihood estimator. 

\item Also, there is no need to assume $\kappa$ to be speaker dependent and average out over all speakers to obtain final estimate of $\kappa$, which is speaker independent. It can be seen later in this section that, the independence property of $\kappa$ is preserved in the Bayesian framework and the large variance of $\kappa$ is reduced in its final estimate.
\end{itemize}
Since, now a days modern Bayesian technique like MCMC \cite{casella1992explaining} is available, in this case Bayesian inference seems to be the obvious choice.

\subsection{Bayesian Estimation Technique}
The model parameters are considered to be random variables for using Bayesian Estimation \cite{hoggintroduction} method. These random variables are assumed to follow a distribution depending on prior knowledge about the parameter and hence these are called 
called the prior distribution. Thereupon the posterior distributions for model parameters are derived using prior distributions and the observed data. The parameters corresponding to the prior distributions are called hyperparameters. The hyperparameters are estimated first in order to obtain better estimate for model parameters. Different kinds of estimates can be obtained using the posterior distribution depending the loss function under consideration.

The affine model for $i$-th reference speaker and a fixed subject speaker is given by
\begin{equation} \label{eq:model_i}
{\bf Y}_i=\alpha_{i} {\bf X}+ \kappa (\alpha_{i}-1) {\bf 1} + {\boldsymbol \epsilon}_{i}
\end{equation}

\noindent where, ${\bf \epsilon}_{i}$ is the error vector. It is assumed to be a multivariate gaussian with zero mean i.e. $ {\bf \epsilon}_{i} \sim \mathcal{N}_r({\bf 0}, \sigma^2 {\bf I_{r,r}})$. Here, r is the length of formant frequency vector of a speaker in the database, ${\bf 0}$ is a vector of zeros of length r and ${\bf I_{r,r}}$ is an identity matrix of size ($r \times r$). From the above Equation, the mean and variance of ${\bf Y}_i$ can be calculated as
\begin{equation} \label{eq:mean}
 {\bf \mu}_i = E({\bf Y}_i) = \alpha_{i} {\bf X}+ \kappa (\alpha_{i}-1) {\bf 1}
\end{equation}
\begin{equation} \label{eq:variance}
 \Sigma_{{\bf Y}_i}=\sigma^2 {\bf I_{r,r}}
\end{equation}

Using Equations (\ref{eq:mean}) and (\ref{eq:variance}) the probability density function of ${\bf Y}_i$ can be written as
\begin{equation} \label{eq:priorK}
f_{y}({\bf Y}_i|\kappa, \alpha_{i}, \sigma)=\frac{1}{\displaystyle (2 \pi \sigma^2)^{r/2}}
e^{\frac{-({\bf Y}_i-{\bf \mu}_i)^T({\bf Y}_i-{\bf \mu}_i)}{2\sigma^2}}
\end{equation}

Now, the prior distribution of $\kappa$ is assumed to be Gaussian i.e. $\kappa \sim \mathcal{N}(a,b^2)$. The conditional posterior distribution of $\kappa$, obtained using Bayes' rule is given below,
\begin{equation} \label{eq:postK}
 f_{\kappa}(\kappa | {\boldsymbol \alpha}, {\bf Y},\sigma)= \frac{1}{\sqrt{2\pi\sigma_{\kappa}^2}}
e^{\frac{-(\kappa-\mu_{\kappa})^2}{2\sigma_{\kappa}^2}}
\end{equation}

\noindent where, $\mu_{\kappa}= \frac{\frac{a}{b^2} + \frac{1}{\sigma^2}\sum_{i=1}^n (\alpha_i-1) ({\bf Y}_i-\alpha_i{\bf X}). {\bf 1} } {\frac{r}{\sigma^2} \sum_{i=1}^n (\alpha_i-1)^2 + \frac{1}{b^2}}$ ,
\hspace{1cm} $\sigma_{\kappa}^2=\frac{1}{\frac{r}{\sigma^2} \sum_{i=1}^n (\alpha_i-1)^2 + \frac{1}{b^2}}$
\vspace{5mm}

\hspace{1cm} ${\bf Y}=({\bf Y_1},{\bf Y_2},\ldots,{\bf Y_n})$ \hspace{3mm} and \hspace{3mm} ${\boldsymbol \alpha} = (\alpha_1, \alpha_2, \ldots, \alpha_n)$ \\

The derivation of Equation (\ref{eq:postK}) is given in \ref{app:a}.
Now, it is very clear from the expression of $\sigma_{\kappa}^2$ that, $\sigma_{\kappa}\leq b$. The equality holds only when $\alpha_i=1$.
So, the variance of the prior distribution of $\kappa$ is reduced in its posterior distribution.

The prior distribution of $\alpha_i$ is also assumed to be Gaussian i.e. $\alpha_i \sim \mathcal{N}(c,d^2)$. So, the conditional posterior distribution of $\alpha_i$ is given by,

\begin{equation} \label{eq:postA}
 f_{\alpha}(\alpha_{i}|\kappa,{\bf Y}_i, \sigma)=\frac{1}{\sqrt{2 \pi \sigma_{\alpha_i}^2}}
 e^{ \displaystyle \frac{-(\alpha_i-\mu_{\alpha_i})^2}{2 \sigma_{\alpha_i}^2} }
\end{equation}

\noindent where, $\mu_{\alpha_i} = \frac{\frac{{\bf X}^T {\bf Y}_i + \kappa ({\bf X} + {\bf Y}_i) {\bf 1} + r \kappa^2}{\sigma^2} + \frac{c}{d^2}}{\frac{1}{d^2} + \frac{{\bf X}^T {\bf X} + 2 \kappa {\bf X}^T {\bf 1} + r \kappa^2}{\sigma^2}}$,
\hspace{1cm} $\sigma_{\alpha_i}^2=\frac{1}{ \frac{1}{d^2} + \frac{{\bf X}^T {\bf X} + 2 \kappa {\bf X}^T {\bf 1} + r \kappa^2}{\sigma^2}}$\\

The derivation of Equation (\ref{eq:postA}) is given in \ref{app:b}.
Now, $\sigma$ is assumed to be uniformly distributed i.e. $\sigma \sim \mathcal{U}(\theta_1,\theta_2)$, its conditional posterior distribution is given by,

\begin{equation} \label{eq:postS}
f_{\sigma}(\sigma | \kappa, {\boldsymbol \alpha},{\bf Y})=\frac{\frac{e^{-\frac{1}{2\sigma^2}\sum_{i=1}^n({\bf Y}_i-{\bf \mu}_i)^T({\bf Y}_i-{\bf \mu}_i)}}{\sigma^{nr}}}
    {\int_{\theta_1}^{\theta_2} \frac{e^{-\frac{1}{2\sigma^2}\sum_{i=1}^n({\bf Y}_i-{\bf \mu}_i)^T({\bf Y}_i-{\bf \mu}_i)}  }{\sigma^{nr}} d\sigma}
\end{equation}

\noindent where, $\sigma \in (\theta_1,\theta_2)$ and ${\bf \mu}_i$ is given in Equation (\ref{eq:mean}).

The denominator in the right hand side of Equation (\ref{eq:postS}) can further be solved as the following,
\begin{equation} \label{eq:postSd}
{\int_{\theta_1}^{\theta_2} \frac{e^{-\frac{\beta}{\sigma^2}}}{\sigma^{nr}} d\sigma} = \frac{1}{2}{\beta}^\frac{nr-1}{2}\gamma(1-\gamma_l-\gamma_u)
\end{equation}

\noindent where, $\beta$, $\gamma$, $\gamma_l$ and $\gamma_u$ are given as follows,
$$\beta = \frac{1}{2}\sum_{i=1}^n({\bf Y}_i-{\bf \mu}_i)^T({\bf Y}_i-{\bf \mu}_i), \hspace{5mm} \gamma = \Gamma\left(\frac{nr-1}{2}\right)$$
$$\gamma_l = \Gamma_{lower}\left(\frac{\beta}{{\theta_2}^2},\frac{nr-1}{2}\right), \hspace{3mm} \gamma_u = \Gamma_{upper}\left(\frac{\beta}{{\theta_1}^2},\frac{nr-1}{2}\right)$$
\vspace{0.5mm}

The derivation of Equations (\ref{eq:postS}) and (\ref{eq:postSd}) is presented in \ref{app:c}.
Now, consider $\Omega$ to be vector of all the model parameters and $\Theta$ to be vector of all hyperparameters i.e.
$$\Omega=(\kappa,\sigma,{\boldsymbol \alpha}), \hspace{5mm} \Theta=(a,b,c,d,\theta_1,\theta_2)$$
The joint distribution of ${\boldsymbol \alpha}$, $\kappa$ and $\sigma$ is given by,
\begin{equation}
 f(\Omega|\Theta,{\bf Y})= f_{1}(\kappa|\Theta_{\kappa})f_{2}(\sigma|\Theta_{\sigma})\prod_{i=1}^{n}f_{y}({\bf Y}_i|\kappa,\alpha_i,\sigma)f_{1}(\alpha_i|\Theta_{\alpha})
 \label{eq:joint}
\end{equation}

\noindent where,
$$\Theta_{\kappa}=(a,b), \hspace{5mm} \Theta_{\alpha}=(c,d), \hspace{5mm} \Theta_{\sigma}=(\theta_1,\theta_2)$$
$$f_{1}(x|m,s)=\frac{1}{\sqrt{2\pi s^2}}e^{\frac{-(x-m)^2}{2s^2}}, \hspace{5mm} f_{2}(\sigma|\theta_1,\theta_2)=\frac{1}{\theta_2-\theta_1}$$
Thus, the joint distribution as well as the conditional posterior distributions are obtained. But, the joint posterior distribution of all parameters are required to obtain the Bayes estimates. Even if the joint posterior distribution is evaluated, it will be very difficult to compute marginal posterior distributions from the joint distribution, because it will involve three dimensional integration. One alternative solution can be to simulate the marginal posterior distributions from the condition posterior distributions.
Gibbs Sampler is used in this work to simulate the distributions. The following section introduces the Gibbs Sampler and discusses the framework of its application on model parameter estimation. It should be noted that the final Bayes estimate depends on the loss function \cite{hoggintroduction} under consideration, e.g. for square error loss function the final estimate is given by mean of the distribution, whereas, any median of the distribution is the final estimate for absolute error loss function. In this paper, square error loss function is considered for all experiments.

\subsection{Model Parameter Estimation using Gibbs Sampler}
Gibbs sampler \cite{casella1992explaining} is a Markov Chain Monte Carlo (MCMC) algorithm which is used to obtain a sequence of observations from a specified probability distribution. In this approach, previous sample value is used to generate next sample in the sequence, which constructs a Markov Chain. The samples of this Markov Chain converge to the required distribution by construction. For example, suppose a bivariate random variable $(x,y)$ is considered, and one wishes to compute the marginals, $p(x)$ and $p(y)$. It is far easier to consider a sequence of conditional distributions, $p(x|y)$ and $p(y|x)$, than it is to obtain the marginals by integration of the joint density $p(x,y)$, i.e.
$$p(x) =\int_{\mathbb D_y} {p(x, y)dy} \hspace{5mm} and \hspace{5mm} p(y) =\int_{\mathbb D_x} {p(x, y)dx}$$
where, ${\mathbb D_x}$ and ${\mathbb D_y}$ are the domains of $x$ and $y$ respectively.
The sampler starts with some initial value $y_0$ for $y$ and obtains $x_1$ by generating a random sample from the conditional distribution $p(x | y = y_0)$. The sampler then uses $x_1$ to generate a new value of $y_1$, drawing from the conditional distribution $p(y| x =x_1)$. The sampler proceeds as follows
$$x_i \sim p(x | y = y_{i-1})\:,\: y_i \sim p(y | x = x_i)$$
Repeating this process N times, generates a Gibbs sequence of length N. First $n(<N)$ terms are rejected to remove the effect of initial guess $y_0$. This Gibbs sequence converges to a stationary distribution that is independent of the starting values, and by construction, this stationary distribution is the target distribution which is being simulated. i.e, $x \sim p(x)$ and $y \sim p(y)$. Also, the expectation of any function $g$ of the random variable $x$ can be approximated in a similar manner. Using the Law of Large Numbers, expected values of $x$ and $g(x)$ can be approximated as follows
\begin{equation}
 \displaystyle \frac{1}{N-n} \displaystyle \sum_{i=n+1}^N x_i \xrightarrow {P} E(x)
\end{equation}
\begin{equation}
 \displaystyle \frac{1}{N-n} \displaystyle \sum_{i=n+1}^N g(x_i) \xrightarrow {P} E[g(x)]
\end{equation}
as $N \rightarrow \infty$ for some large $n$. So, the estimates are obtained by taking average of the generated samples.

\subsubsection{Algorithm for Model Parameter Estimation using Gibbs Sampler}
The steps involved in calculation of the expected value of parameters of the affine model using Gibbs Sampler are enumerated in Algorithm 1.
\begin{algorithm}
\label{alg1}
\caption{Parameter Estimation using Gibbs Sampler}
\begin{algorithmic}[1]
\State {\bf {Posterior Distributions}} : The posterior distributions of $\kappa \sim f_{\kappa}( \kappa | \alpha_1, \ldots, \alpha_n , \sigma, {\bf Y})$, $\alpha_i \sim f_{\alpha}(\alpha_i| \kappa, \sigma, {\bf Y}_i)$ and $\sigma \sim f_{\sigma}(\sigma | \alpha_1, \ldots, \alpha_n, \kappa, {\bf Y})$ are considered as given in Equations (\ref{eq:postK}), (\ref{eq:postA}) and (\ref{eq:postS}) respectively.
\State {\bf{Initial Guess}} : Initial guess is made for $\kappa$ and $\sigma$ as $\kappa=\kappa^{(0)}$ and $\sigma=\sigma^{(0)}$
\State {\bf{Sampling of $\alpha$}} : The j-th sample of $\alpha_i$, i.e. $\alpha_i^{(j)}$ is generated from its posterior distribution, for $i=1,2,\ldots,n$, \\ $\alpha_i^{(j)}  \sim f_{\alpha}(\alpha_i| \kappa^{(j-1)}, \sigma^{(j-1)}, {\bf Y}_i)$
\State {\bf{Sampling of $\kappa$}} : The j-th sample of $\kappa$, i.e. $\kappa^{(j)}$ is generated from its posterior distribution, \\ $\kappa^{(j)}  \sim f_{\kappa}(\kappa | \alpha_1^{(j)},\ldots, \alpha_n^{(j)}, \sigma^{(j-1)}, {\bf Y})$
\State {\bf{Sampling of $\sigma$}} : The j-th sample of $\sigma$, i.e. $\sigma^{(j)}$ is generated from its posterior distribution, \\ $\sigma^{(j)}  \sim f_{\sigma}(\sigma |\alpha_1^{(j)},\ldots,\alpha_n^{(j)}, \kappa^{(j)}, {\bf Y})$ \cite{truncated}
\State {\bf{Iteration}} :  The steps 3, 4 and 5 are repeated for $j = 1, 2, \ldots, M$, where, $M$ is the number of iterations.
%
\end{algorithmic}
\end{algorithm}

Finally, the expected values of $\alpha_i$, $\kappa$ and $\sigma$ are calculated as follows,
\begin{equation}
E(\alpha_i) = \displaystyle \frac{1}{M-m} \displaystyle \sum_{j=m+1}^M \alpha_i^{(j)}
\end{equation}
\begin{equation}
E(\kappa) = \displaystyle \frac{1}{M-m} \displaystyle \sum_{j=m+1}^M \kappa^{(j)}
\end{equation}
\begin{equation}
E(\sigma) = \displaystyle \frac{1}{M-m} \displaystyle \sum_{j=m+1}^M \sigma^{(j)}
\end{equation}
for large $m$ and $M$ such that, $ m < M $. Here, $m$ is burn-in period.

\subsubsection{Variation of Model Parameters with respect to Hyperparameters}
\begin{figure}[h]
\begin{minipage}[b]{0.48\linewidth}
  \centering
  \centerline{\includegraphics[scale=0.4]{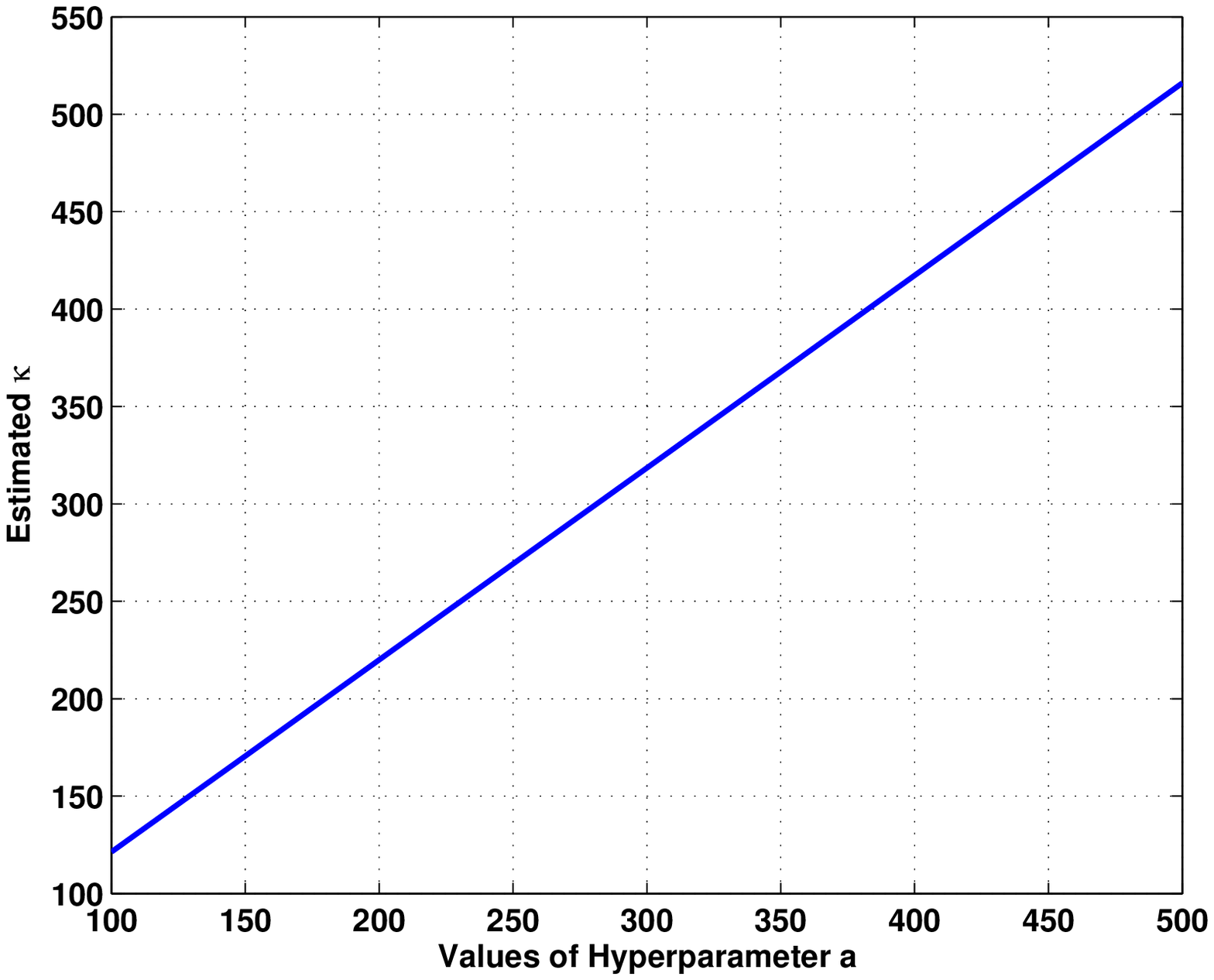}}
  \label{fig:kappa_b15}
  \centerline{Estimated $\kappa$ vs a}
\end{minipage}
\begin{minipage}[b]{0.48\linewidth}
  \centering
  \centerline{\includegraphics[scale=0.4]{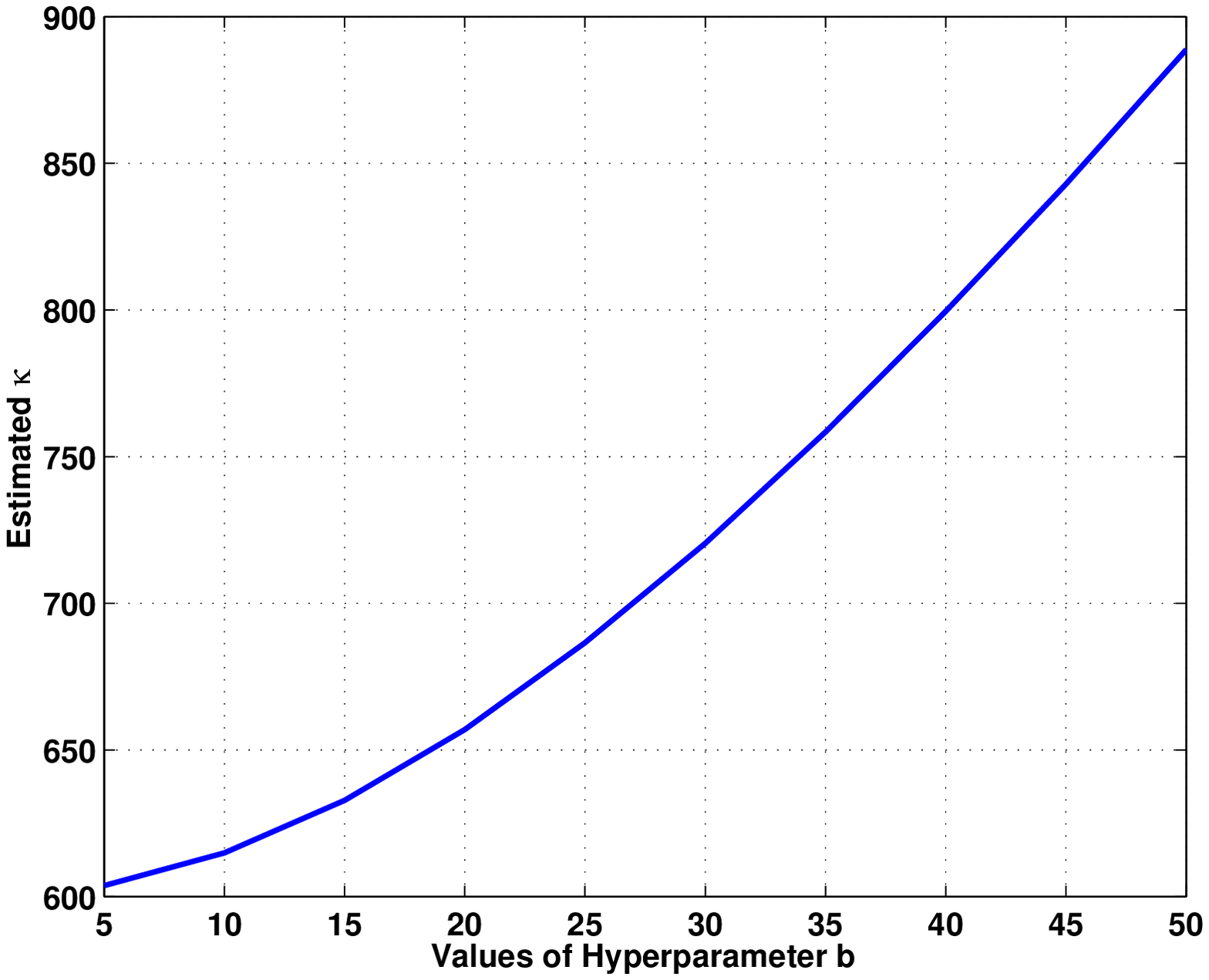}}
  \label{fig:kappa_b50}
  \centerline{Estimated $\kappa$ vs b}
\end{minipage}
\caption{Illustration of variation of estimated values of $\kappa$ with varying values of hyperparameters}
\label{fig:simulatedK}
\end{figure}

In order to observe the effects of hyperparameters on the estimated values of model parameters, simulations are performed by varying the value of one hyperparameter while keeping others constant. In simulation, the number of Gibbs runs are 2000 and the burn in period is 1500, i.e. only last 500 values are taken into consideration among 2000 estimated values to get final estimate of $\kappa$. The simulation results are shown in Figure \ref{fig:simulatedK}. The diagrams presented in Figure \ref{fig:simulatedK} show the variation in estimated value of $\kappa$ with hyperparameters $a$ and $b$ respectively. The plots indicate that, estimates of $\kappa$ are highly dependent on both $a$ and $b$. Similar experiments show large dependency of estimates of $\kappa$, ${\boldsymbol \alpha}$ and $\sigma$ on other hyperparameters also. From these observations it can be concluded that, hyperparameters need to estimated first to get better estimate of the model parameters. The estimation of hyperparameters is discussed in the ensuing section.

\subsection{Maximum Likelihood Estimation of Hyperparameters}
Hyperparameters of the affine model are estimated using Likelihood Maximization technique. The likelihood function is calculated by multiplying distributions of all the parameters. This function contains both, model parameters and hyperparameters. The model parameters are integrated out to obtain a likelihood function, dependent only on the hyperparameters. The required estimate of the hyperparameters are values for which the likelihood function is maximized.

Under the assumption, $\kappa \sim \mathcal{N}( a, b^2)$, $\alpha_i \sim \mathcal{N}(c, d^2)$ and $\sigma \sim \mathcal{U}(\theta_1, \theta_2)$ and ${\bf \epsilon}_{i} \sim \mathcal{N}_r({\bf 0},\sigma^2 {\bf I_{r,r}})$, the likelihood function is given by,
\begin{equation}
 L(\Theta|\Omega,{\bf Y})=f_{1}(\kappa|\Theta_{\kappa})f_{2}(\sigma|\Theta_{\sigma})\prod_{i=1}^{n}f_{y}({\bf Y}_i|\kappa,\alpha_i,\sigma)f_{1}(\alpha_i|\Theta_{\alpha})
\label{eq:likeli}
\end{equation}

\noindent where, $\Theta,\Omega,{\bf Y},f_{1}(\kappa|\Theta_{\kappa}),f_{2}(\sigma|\Theta_{\sigma})$ and $f_{1}(\alpha_i|\Theta_{\alpha})$ are defined in Section \ref{sec:bayesian}.
The likelihood function in Equation (\ref{eq:likeli}) contains terms of $a$, $b$, $c$, $d$, $\theta_1$, $\theta_2$, $\kappa$, $\alpha_1, \ldots, \alpha_n$, and $\sigma$. But the required likelihood function should be a function of hyperparameters and it should not contain model parameters. In order to achieve this, the model parameters $\kappa$, $\alpha_1, \ldots, \alpha_n$ and $\sigma$ are integrated out from Equation (\ref{eq:likeli}) to obtain the integrated likelihood ($IL(\Theta)$) function as follows, 
\begin{equation} \label{eq:IL1}
IL(\Theta) = \int_{\theta_1}^{\theta_2}\int_{-\infty}^{\infty} \frac{f(\kappa,\:\sigma,\:c,\:d)}{(\theta_2 - \theta_1){\sqrt{2 \pi b^2}}} e^{ \frac{-(\kappa-a)^2}{2 b^2} }  d\kappa d\sigma
\end{equation}

\noindent where, $f(\kappa,\:\sigma,\:c,\:d)$ denotes a part of the likelihood function which is obtained after integrating out all the $\alpha_i$'s.
\begin{equation} \label{eq:fkscd1}
 f(\kappa,\:\sigma,\:c,\:d) = \prod_{i=1}^n \int_{-\infty}^{\infty}\frac{e^{\frac{-(\alpha_i-c)^2}{2 d^2}+ \frac{-({\bf Y}_i-{\bf \mu}_i)^T({\bf Y}_i-{\bf \mu}_i)}{2 \sigma^2}}}
 {\sqrt{2 \pi d^2} (2 \pi \sigma^2)^{r/2}}  d \alpha_i
\end{equation}

\noindent which simplifies to the following expression,
%

\begin{equation} \label{eq:fkscd2}
f(\kappa,\:\sigma,\:c,\:d) = \frac{A_{\alpha_i}^{\frac{-n}{2}}}{2^{\frac{n(r+1)}{2}}\pi^{\frac{nr}{2}} d^n \sigma^{nr}} e^{\sum_{i=1}^{n}\left( \frac{B_{\alpha_i}^2}{A_{\alpha_i}} - C_{\alpha_i}\right)}
\end{equation}

where, $ A_{\alpha_i} = \frac{1}{2d^2} + \frac{{\bf X}^T {\bf X} + 2 \kappa ({\bf X}^T {\bf 1}) + r \kappa^2}{2\sigma^2} $
\vspace{2mm}

\hspace{10.5mm}$ B_{\alpha_i} = \frac{{\bf X}^T {\bf Y}_i + \kappa ({\bf X} + {\bf Y}_i)^T {\bf 1} + r \kappa^2}{2\sigma^2} + \frac{c}{2d^2} $
\vspace{2mm}

\hspace{10.5mm}$ C_{\alpha_i} = \frac{{\bf Y}_i^T {\bf Y}_i + 2\kappa ({\bf Y}_i^T {\bf 1}) + r \kappa^2}{2\sigma^2} + \frac{c^2}{2d^2} $
\vspace{2mm}

Now using Equation (\ref{eq:fkscd2}), Equation (\ref{eq:IL1}) can be written as,
\begin{equation} \label{eq:IL2}
IL(\Theta) = \int_{\theta_1}^{\theta_2}\int_{-\infty}^{\infty} \frac{A_{\alpha_i}^{\frac{-n}{2}} e^{\frac{-(\kappa - a)^2}{2b^2}\sum_{i=1}^{n}\left( \frac{B_{\alpha_i}^2}{A_{\alpha_i}} - C_{\alpha_i}\right)} }{2^{\frac{n(r+1)+1}{2}}\pi^{\frac{(nr+1)}{2}} d^n \sigma^{nr} (\theta_2 - \theta_1)}\: d\kappa d\sigma
\end{equation}

The derivation of Equations (\ref{eq:IL1}) through (\ref{eq:IL2}) is given in \ref{app:d}. The values of $a,\:b,\:c,\:d,\:\theta_1$ and $\theta_2$ for which the integrated likelihood displayed in Equation (\ref{eq:IL2}) attains maximum value for given ${\bf Y}$, irrespective of $\kappa$, $\alpha_1, \ldots, \alpha_n$ and $\sigma$, are the final estimates of hyperparameters. Note that, the constant in Equation (\ref{eq:IL2}) can be ignored, because it does not affect maximization. So, the integrated likelihood can be written as,
\begin{equation} \label{eq:IL3}
 IL(\Theta)=\int_{\theta_1}^{\theta_2}\int_{-\infty}^{\infty} \frac{A_{\alpha_i}^{\frac{-n}{2}}} { d^n \sigma^{nr} (\theta_2 - \theta_1)} e^{\frac{-(\kappa - a)^2}{2b^2}\sum_{i=1}^{n}\left( \frac{B_{\alpha_i}^2}{A_{\alpha_i}} - C_{\alpha_i}\right)}\: d\kappa d\sigma
\end{equation}

\noindent Now, instead of calculating $IL(\Theta)$, $\log(IL(\Theta))$ is considered for optimization, because $\log$ is a concave function and it will not affect the maximization. The optimum value of parameters are obtained by maximizing $\log(IL(\Theta))$.

\begin{figure}
\begin{minipage}[b]{0.48\linewidth}
  \centering
  \centerline{\includegraphics[scale=0.4]{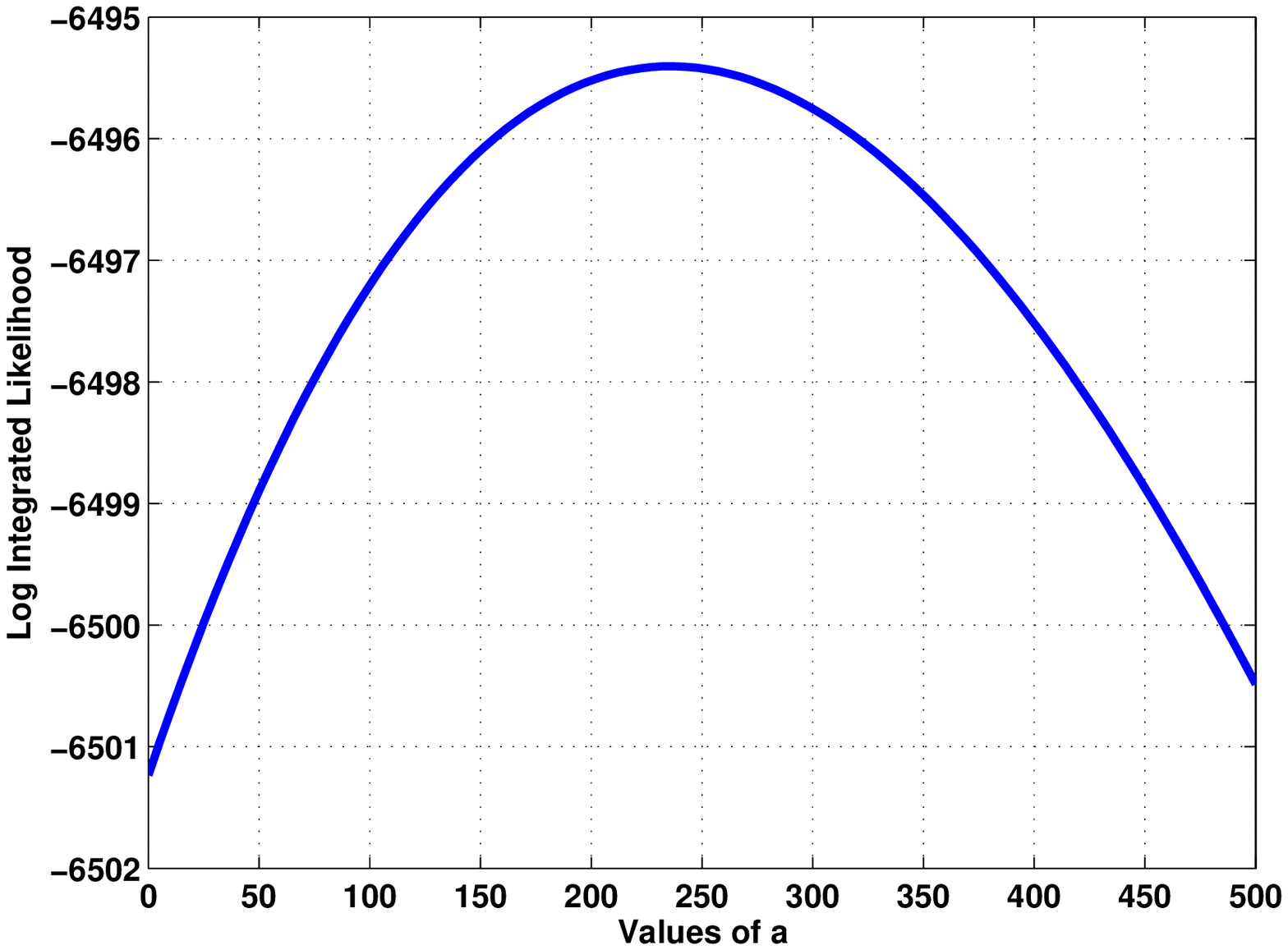}}
\label{fig:logil_a}
  \centerline{(a) $\log(IL(\Theta)) \hspace{1mm}vs\hspace{1mm} a$}
\end{minipage}
\begin{minipage}[b]{0.48\linewidth}
  \centering
  \centerline{\includegraphics[scale=0.4]{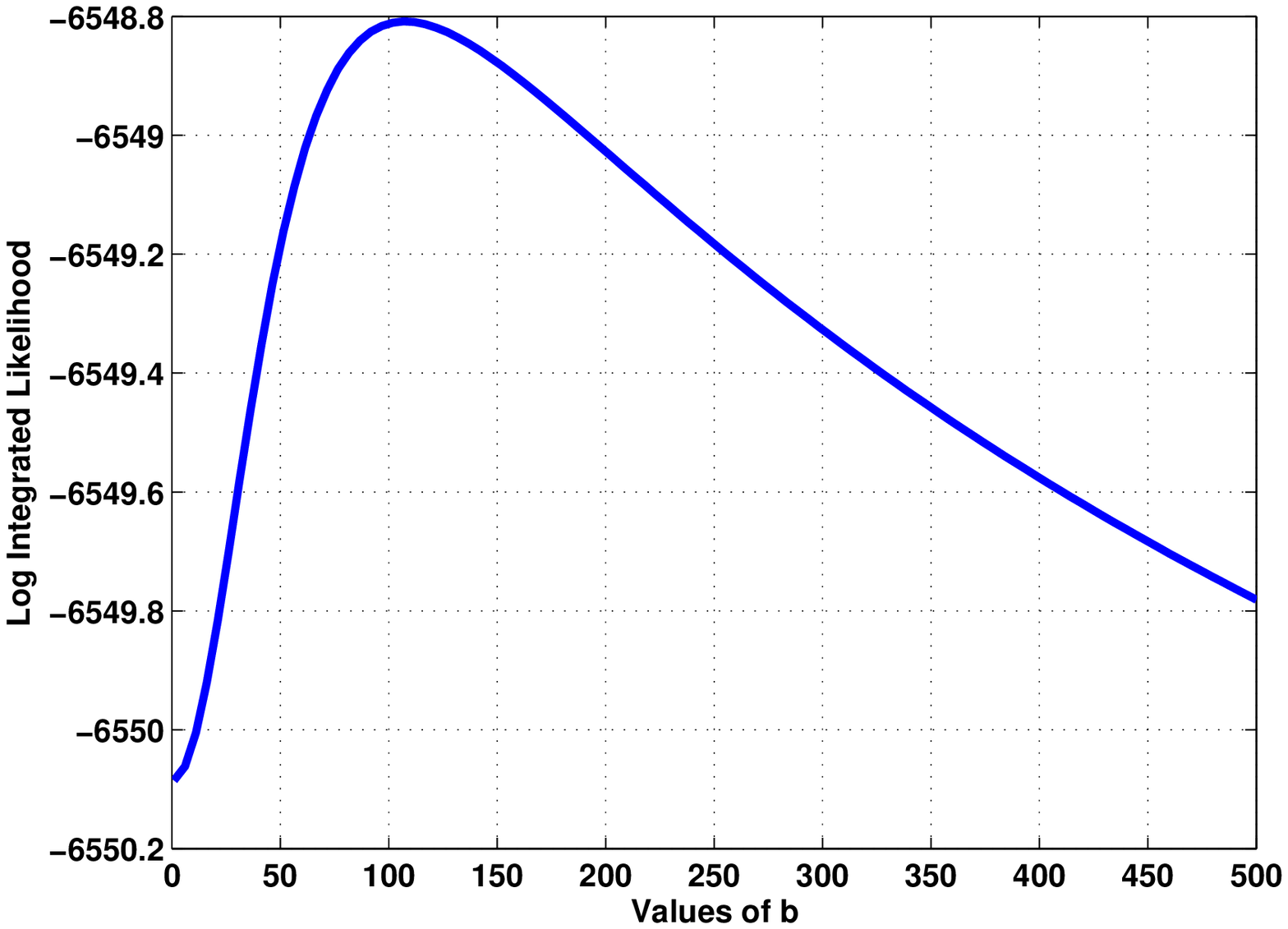}}
\label{fig:logil_b}
  \centerline{(b) $\log(IL(\Theta)) \hspace{1mm}vs\hspace{1mm} b$}
\end{minipage}
\begin{minipage}[b]{0.48\linewidth}
  \centering
  \centerline{\includegraphics[scale=0.4]{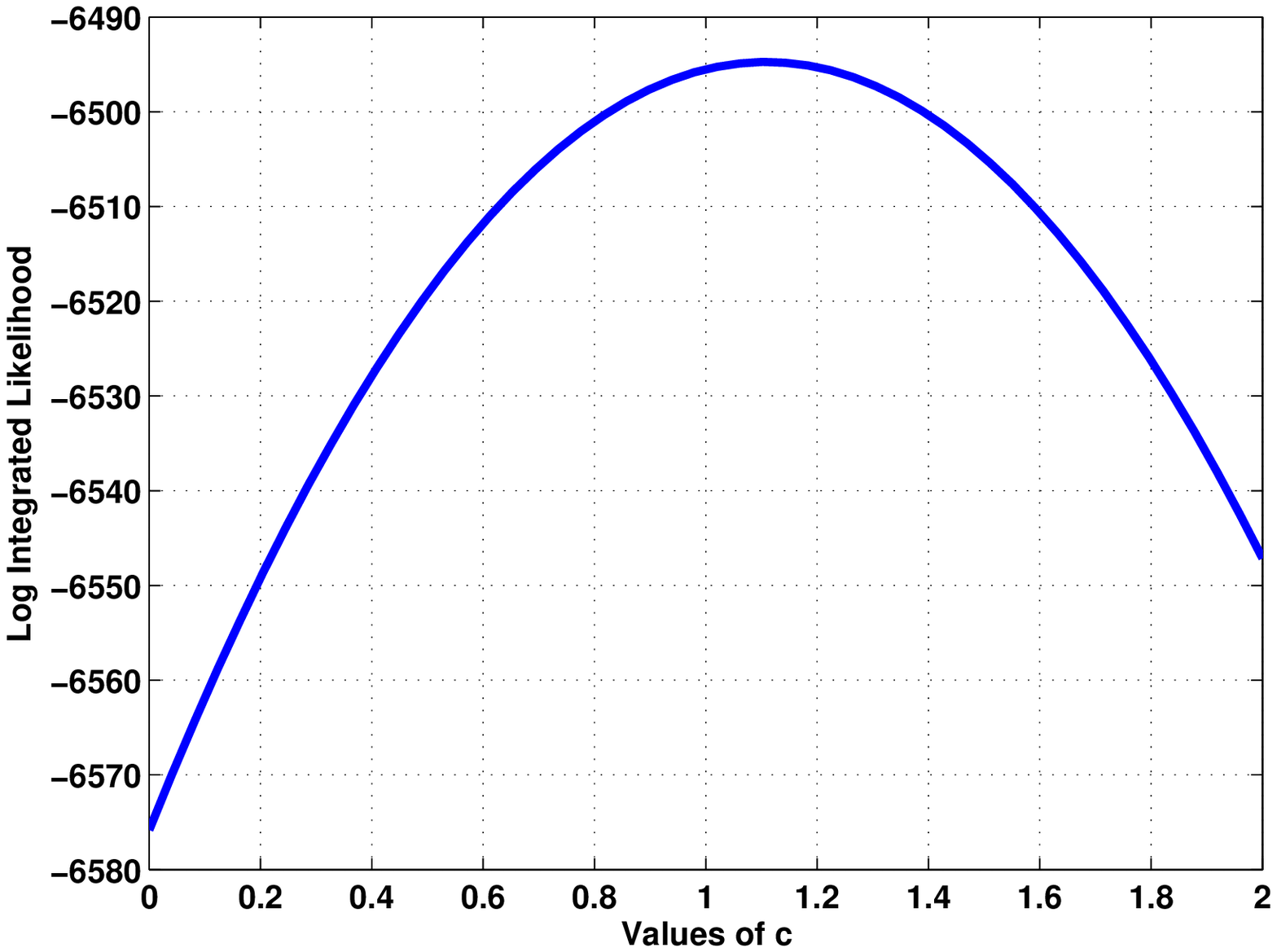}}
\label{fig:logil_c}
  \centerline{(c) $\log(IL(\Theta)) \hspace{1mm}vs\hspace{1mm} c$}
\end{minipage}
\begin{minipage}[b]{0.48\linewidth}
  \centering
  \centerline{\includegraphics[scale=0.4]{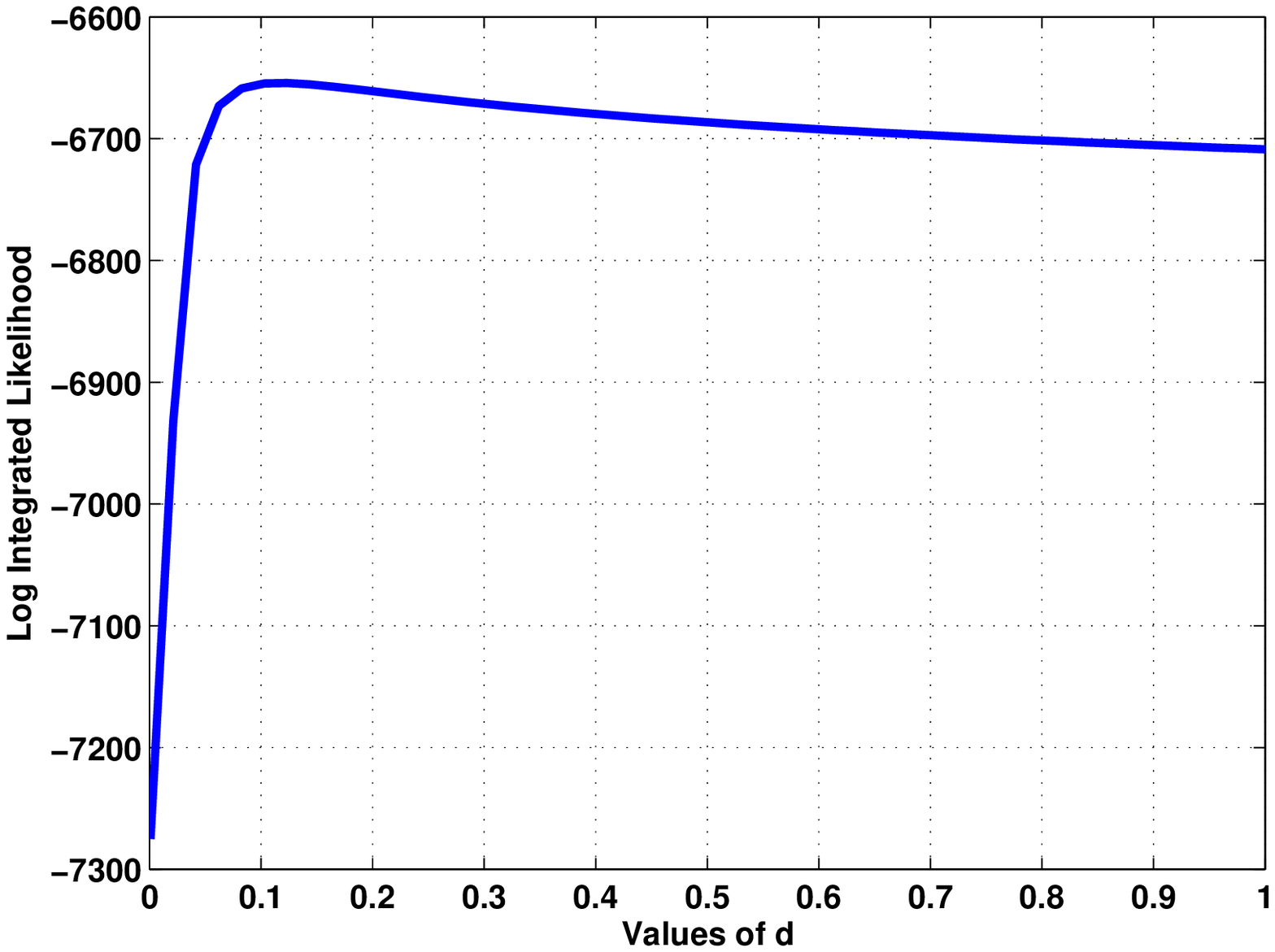}}
\label{fig:logil_d}
  \centerline{(d) $\log(IL(\Theta)) \hspace{1mm}vs\hspace{1mm} d$}
\end{minipage}
\begin{minipage}[b]{0.48\linewidth}
  \centering
  \centerline{\includegraphics[scale=0.4]{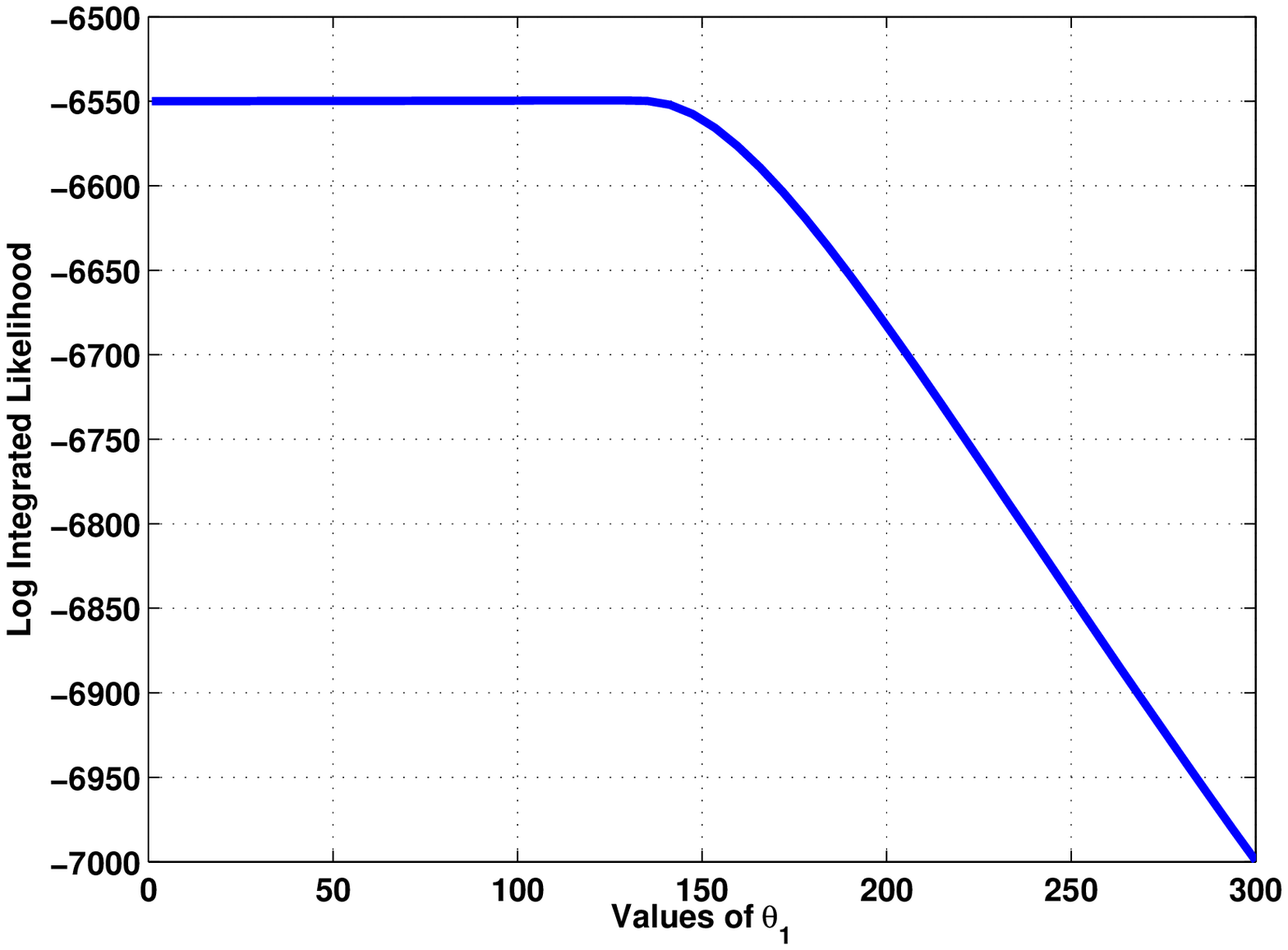}}
\label{fig:logil_t1}
  \centerline{(e) $\log(IL(\Theta)) \hspace{1mm}vs\hspace{1mm} \theta_1$}
\end{minipage}\hspace{3mm}
\begin{minipage}[b]{0.48\linewidth}
  \centering
  \centerline{\includegraphics[scale=0.4]{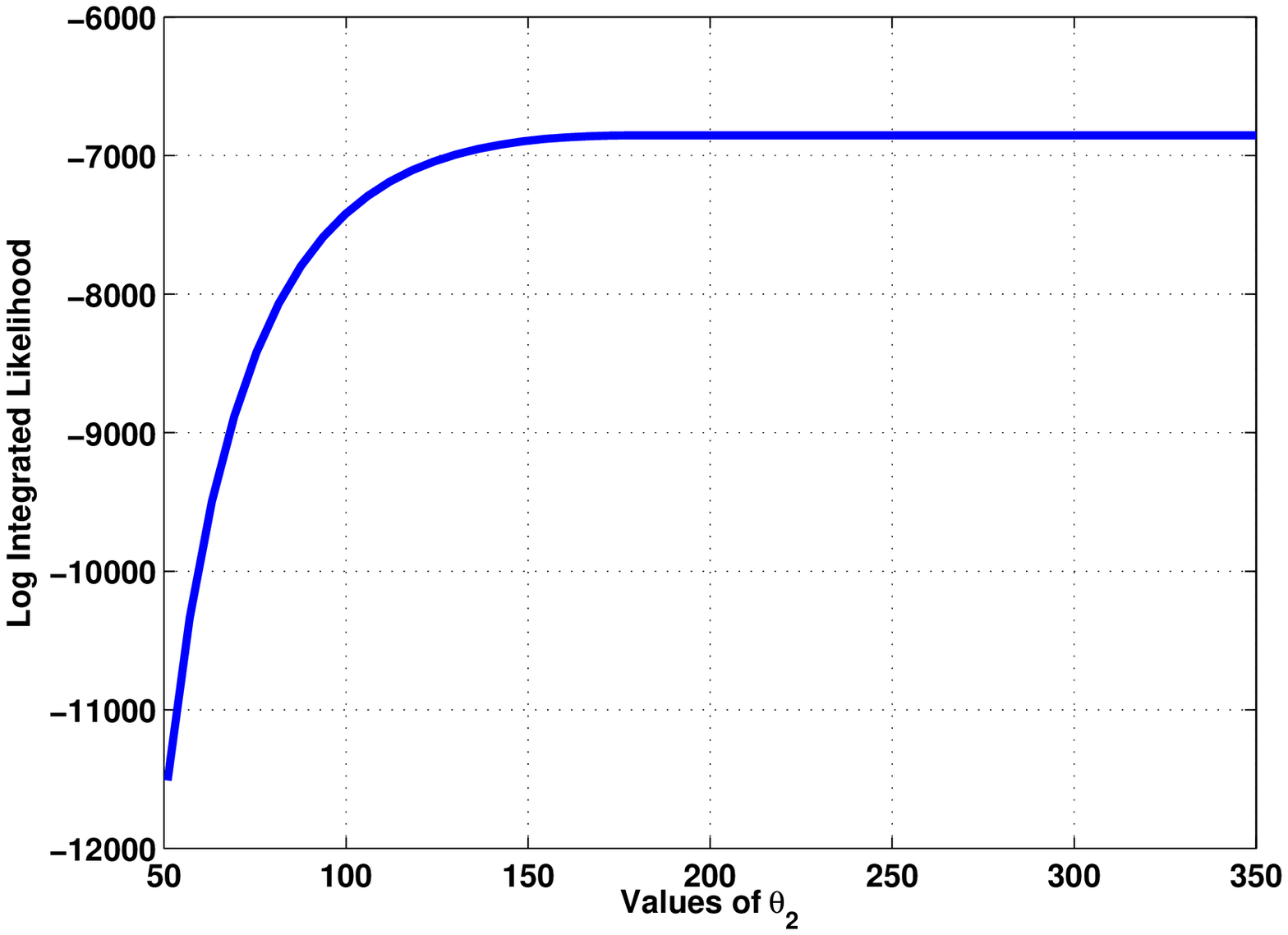}} 
\label{fig:logil_t2}
  \centerline{(f) $\log(IL(\Theta)) \hspace{1mm}vs\hspace{1mm} \theta_2$}
\end{minipage}
\caption{Figures illustrating the variation of Log Integrated Likelihood ($\log(IL(\Theta))$) with respect to $a$, $b$, $c$, $d$, $\theta_1$ and $\theta_2$ respectively, varying one at a time keeping others constant}
\label{fig:logil}
\end{figure}
Therefore, the task remains to maximize $\log(IL(\Theta))$ with respect to $a$, $b$, $c$, $d$, $\theta_1$ and $\theta_2$. Prior to carrying out the maximization of $\log(IL(\Theta))$, nature of the objective function needs to be examined. The variation of $\log(IL(\Theta))$ against each of the six hyperparameters are shown in Figure \ref{fig:logil}. 
A careful observation of these plots indicate that, $\log(IL(\Theta))$ is uni-modal with respect to all hyperparameters. Hence, $\log(IL(\Theta))$ can be maximized. A joint maximization is needed over these six variables to obtain the required hyperparameters. Interior-point algorithm \cite{karmarkar1984new} has been used to solve this optimization problem. This is a special kind of linear programming algorithm in which the optimal solution is reached by traversing the interior of the feasible region. Here, the feasible region is the set of all possible points of an optimization problem that satisfy the problem's constraints.

\subsection{Algorithm to Estimate Model Parameter using Bayesian Estimation Technique}
The steps for calculating the affine model parameters using Bayesian Estimation are enumerated in Algorithm 3. This algorithm improves performance of the recognizer by a significant amount. The improvement in recognition accuracy is shown in the following section with the help of two kinds of recognition experiment. Although the number of parameters were kept in check as described in Section \ref{sec:model}, the computational cost was higher than the conventional linear VTLN method \cite{lee1996speaker}. Thus, both normalization models will find its application depending on the problem under consideration.

\begin{algorithm}[!h]
\label{alg1}
\caption{Estimation of Parameters $\alpha$ and $\kappa$ using Bayesian Estimation Method}
\begin{algorithmic}[1]
\State {\bf {Formant Frequency Vectors}} : Formant frequencies are extracted \cite{snell1993formant} from all utterances to construct fromant frequency vector for subject speaker ({\bf X}) and reference speaker ({\bf Y}).
\State {\bf{Integrated likelihood function}} : The integrated likelihood function is calculated as shown in Equation (\ref{eq:IL3}).
\State {\bf{Estimation of Hyperparameters}} : The hyperparameters $a, b, c, d, \theta_1, \theta_2$ are calculated by optimizing Equation (\ref{eq:IL3}) using interior point algorithm.
\State {\bf{Estimation of $\alpha$ and $\kappa$}} : Using the estimates of hyperparameters, the affine model parameters $\alpha$ and $\kappa$ are calculated using Algorithm2.
\end{algorithmic}
\end{algorithm}

%% file: experiments.tex
\section{Performance Evaluation}
\label{sec:PerEv}
In order to evaluate performance of the Bayesian approach for speaker normalization proposed in this paper, two different experiments are conducted namely, vowel recognition experiment and speech recognition experiment. Vowel recognition experiments are conducted to validate the whole set-up of speaker normalization, because it is based on formant frequencies. Afterwards, speech recognition experiments are carried out to demonstrate the scope of the proposed approach for real-life applications. 

\subsection{Experiments on Vowel Recognition}
The vowel recognition experiments are performed using a Mahalanobis distance \cite{mahalanobis} based vowel recognizer. A brief description is presented on the recognizer as well as the databases being used for vowel recognition and the experimental set-up is discussed. Thereafter the experiments are conducted and recognition performances are shown in terms of recognition accuracy. Note that, the experiments on vowels are solely based on their formant frequencies and not on any other acoustic information from the utterances of vowels.

\subsubsection{Formant based Vowel Recognizer}
Formant frequency vectors corresponding to each speaker are needed to implement the normalization scheme. The formant frequency vector for a particular speaker is constructed by concatenating formants of all vowels spoken by that speaker. Subsequently, the methods discussed in Sections \ref{sec:speaker_norm} and \ref{sec:bayesian} are used to compute the normalization parameters for each speaker. The estimation of normalization parameters is followed by its application on formant frequencies of the database to obtain normalized frequencies using Equation (\ref{eq:estimate}), as shown in Section \ref{sec:speaker_norm}. First three formant frequencies corresponding to a vowel are considered for recognition. The reference speakers' database contains different vowels spoken by various speakers, which constitutes various instances for each vowel. Finally for testing a vowel utterance, its formant frequencies are extracted and its Mahalanobis distance is computed from each vowel group present in the reference speakers' database. The vowel corresponding to the minimum distance is the recognized vowel.

\subsubsection{Vowel Databases}
The following two databases are used for vowel recognition experiments.
\vspace{-1.5mm}
\begin{itemize}
\item {\it Peterson $\&$ Barney Database (PnB):}
There are a total of 76 speakers (33 Males, 28 Females and 15 Children) in PnB database \cite{peterson1952control}. The utterances were recorded using magnetic tape recorder. For the recordings, a list of 10 monosyllabic words were prepared each starting with `h' and ending with `d'. Speakers were given a list of words before recording. The order in the lists were randomized, and each speaker was asked to pronounce words using two different lists. Randomizing the list avoids the practice effects of the speakers. This database consisted of utterances of 10 vowels (/aa/, /ae/, /ah/, /ao/, /eh/, /er/, /ih/, /iy/, /uh/, /uw/), which are uttered twice by each of the speakers. Alternately, the PnB database can be considered to be having 152 speakers (66 Males, 56 Females and 30 Children), with each of them having uttered 10 vowels once.

\item {\it Hillenbrand Database (Hil):}
The Hil database \cite{hillenbrand1995acoustic} effectively consists of a total of 98 speakers (37 Males, 33 Females and 28 Children). Here each of the speakers have uttered only once, each of the 12 vowels (/ae/, /ah/, /aw/, /eh/, /ei/, /er/, /ih/, /iy/, /oa/, /oo/, /uh/, /uw/). These vowels are extracted from 12 monosyllabic words starting with `h' and ending with `d', which were uttered by those speakers. There were some more speakers in this database, but they are not considered for our experiments as some of the formants corresponding to those speakers are marked zero. The formants were marked zero because the authors were unable to calculate them.
The aforementioned databases are available in \cite{PnBdata} and \cite{Hildata} respectively.
\end{itemize}

\subsubsection{Experimental Conditions}
Each speaker in both databases are characterized by using first three formant frequencies ($F_1, F_2, F_3$) of each vowel. Formant frequency vector corresponding to a speaker is constructed by concatenating the formant frequencies of all vowels spoken by that speaker. Since there are three formant frequencies for each vowel, the dimension of the formant frequency vector corresponding to each speaker will be 30 for PnB database (corresponding to 10 vowels) and 36 for Hil database (corresponding to 12 vowels).

In both databases mentioned above there are three categories of speakers: male, female and child. Thus, the normalization parameters, $\kappa$ (speaker independent) and $\alpha$ (speaker dependent) are calculated using all combinations of these three categories, e.g. for male speakers, three different kinds of normalization parameters are obtained by taking male speaker as subject and alternately considering male, female and child speaker as reference. This is followed by normalization of the databases using estimated parameters. Hence, for each kind of speaker there will be three classes of normalized frequencies corresponding to different categories of reference speakers used. In total, there will be 9 different combinations of subject and reference speaker as MM, MF, MC, FM, FF, FC, CM, CF and CC, where M, F, C correspond to Male, Female and Child speaker respectively.

\begin{figure}[!h]
\begin{minipage}{0.33\textwidth}
  \centering
  \centerline{\includegraphics[width=\linewidth,height=4.2cm]{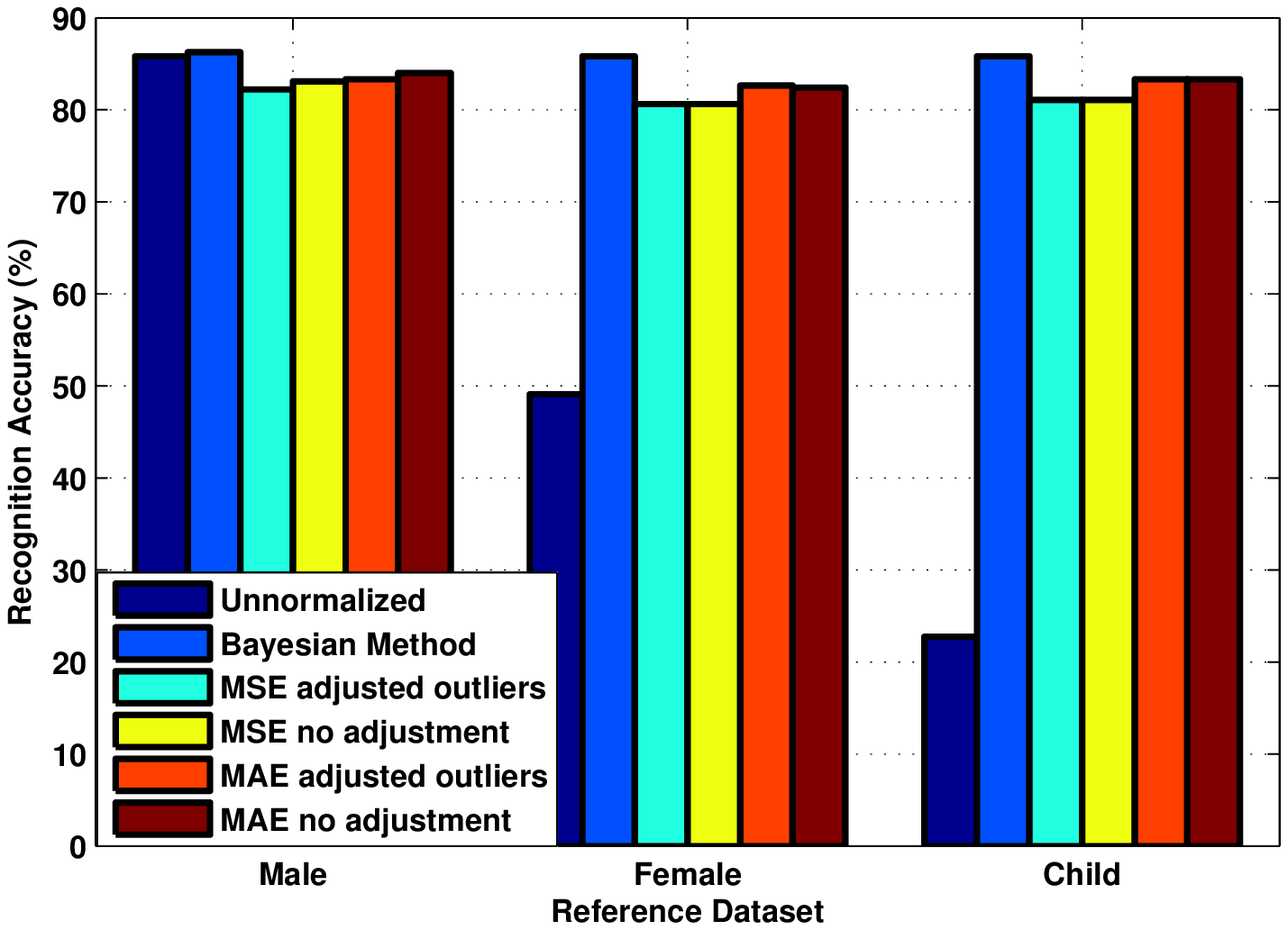}}
  \centerline{(a) Male Subject}
\end{minipage}\hfill
\begin{minipage}{0.33\textwidth}
  \centering
  \centerline{\includegraphics[width=\linewidth,height=4.2cm]{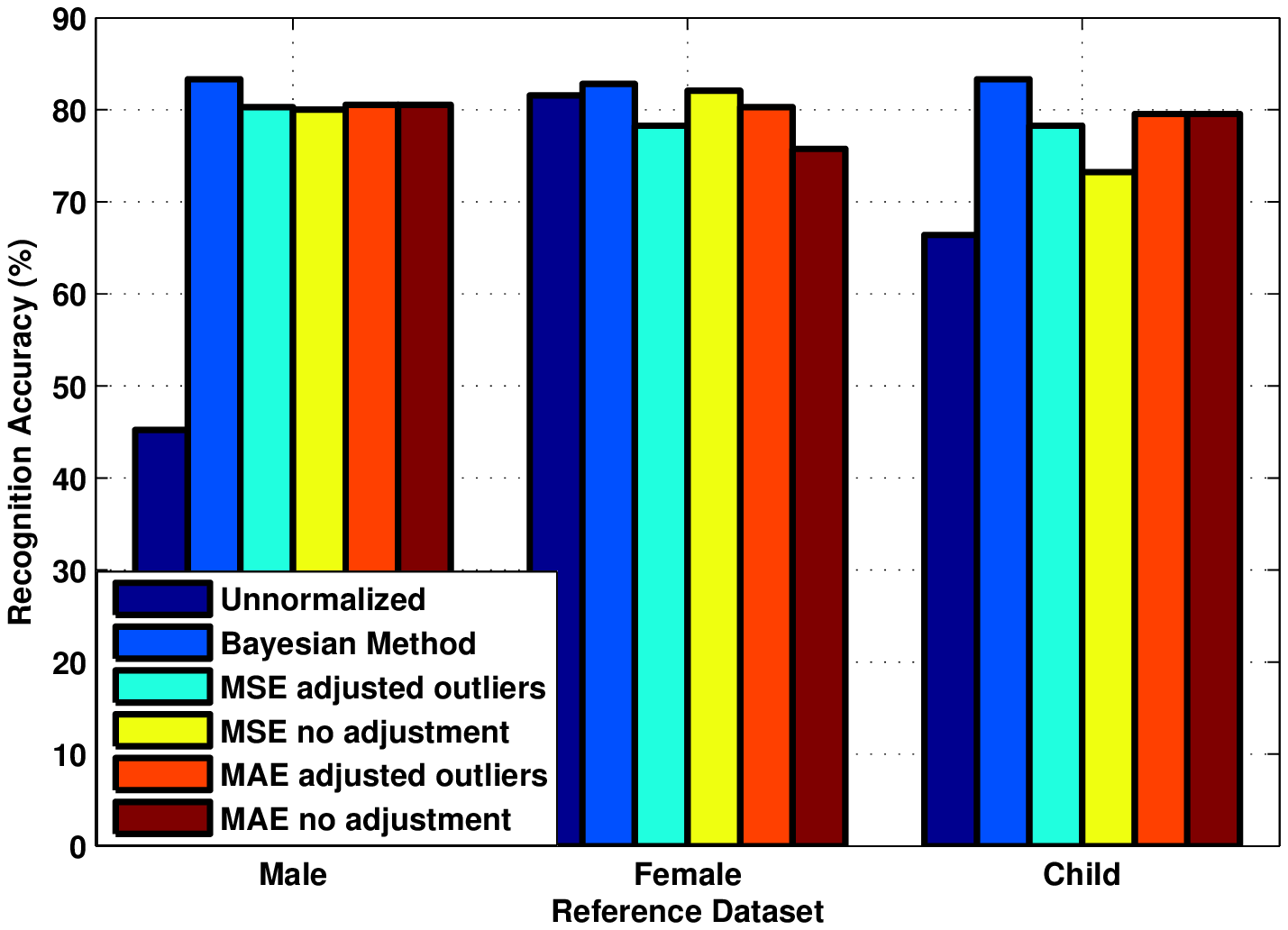}}
  \centerline{(b) Female Subject}
\end{minipage}\hfill
\begin{minipage}{0.33\textwidth}
  \centering
  \centerline{\includegraphics[width=\linewidth,height=4.2cm]{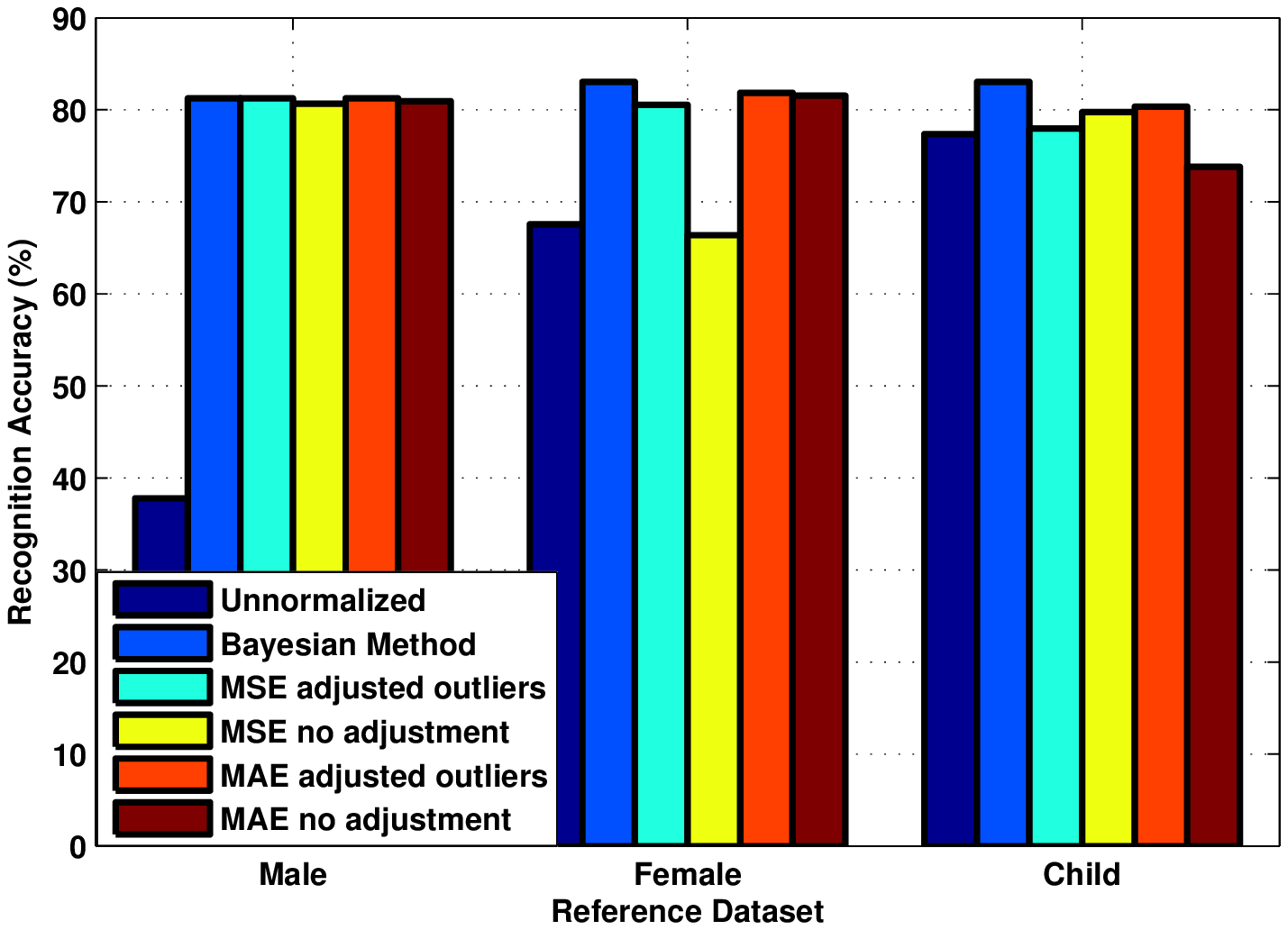}}
  \centerline{(c) Child Subject}
\end{minipage}
\caption{Bar diagrams showing vowel recognition performance on Hil database in terms of Recognition Accuracy for baseline case and using normalization}
\label{hil_nr_nt}
\end{figure}

\begin{figure}[!h]
\begin{minipage}{0.33\textwidth}
  \centering
  \centerline{\includegraphics[width=\linewidth,height=4.2cm]{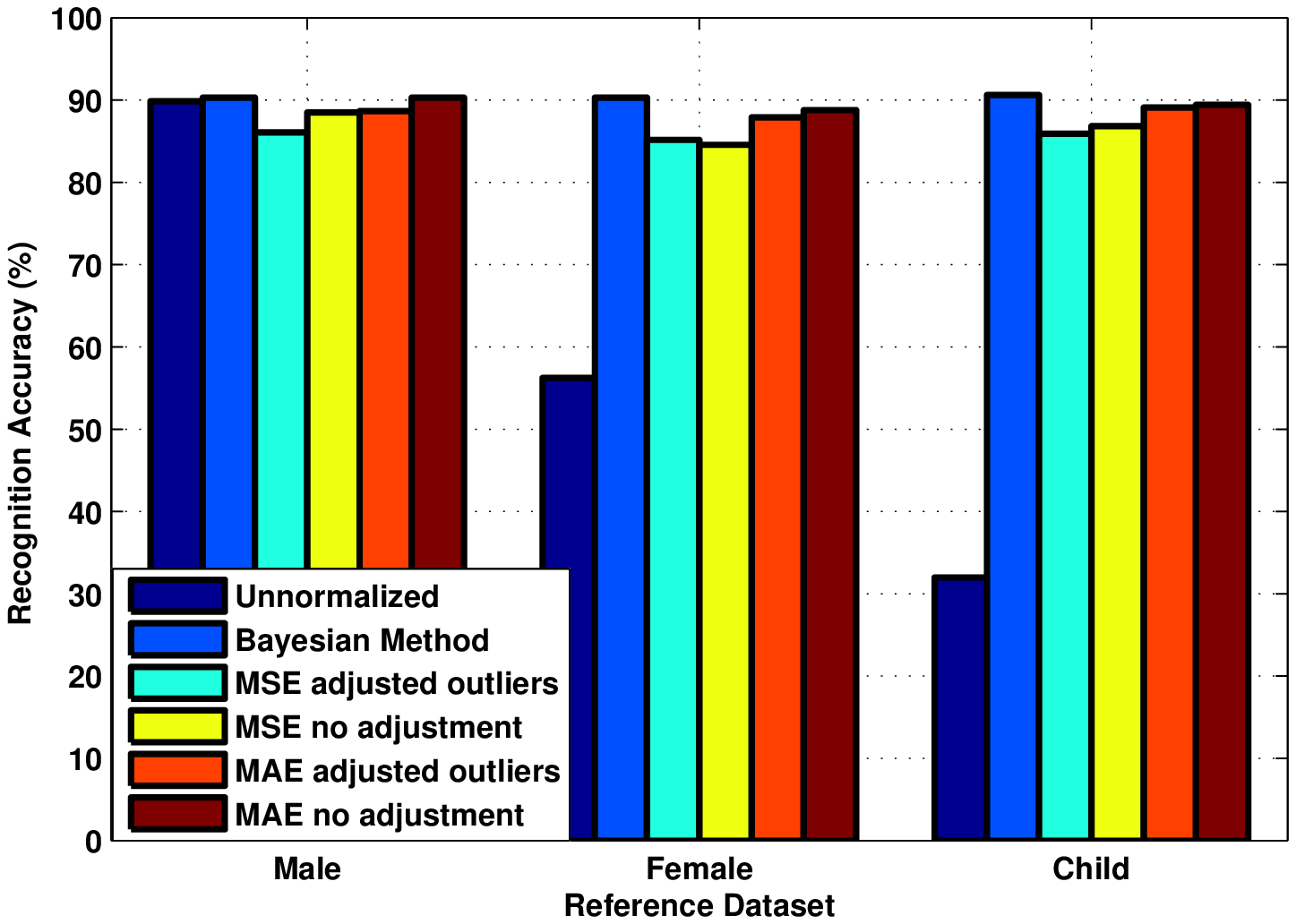}}
  \centerline{(a) Male Subject}
\end{minipage}\hfill
\begin{minipage}{0.33\textwidth}
  \centering
  \centerline{\includegraphics[width=\linewidth,height=4.2cm]{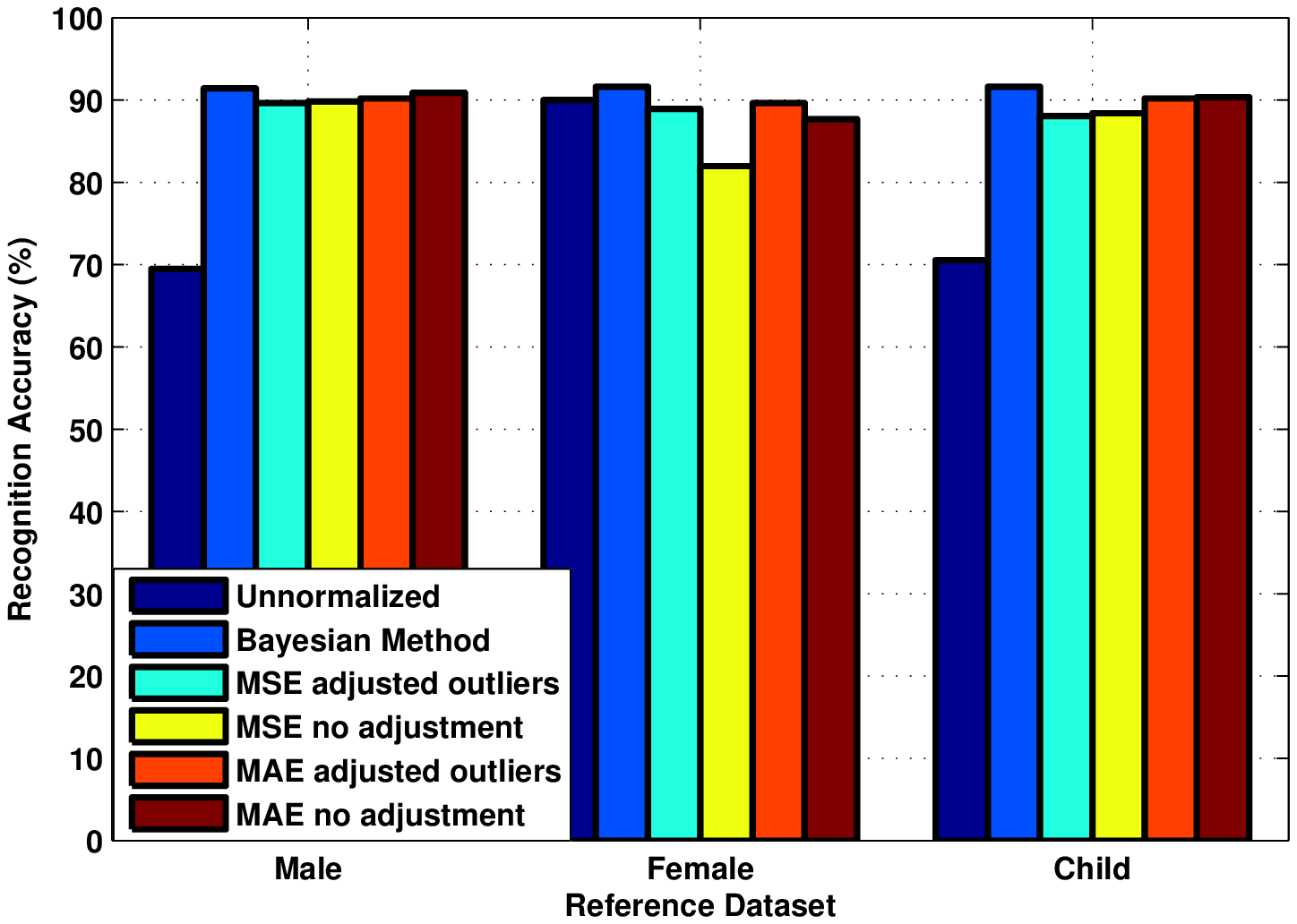}}
  \centerline{(b) Female Subject}
\end{minipage}\hfill
\begin{minipage}{0.33\textwidth}
  \centering
  \centerline{\includegraphics[width=\linewidth,height=4.2cm]{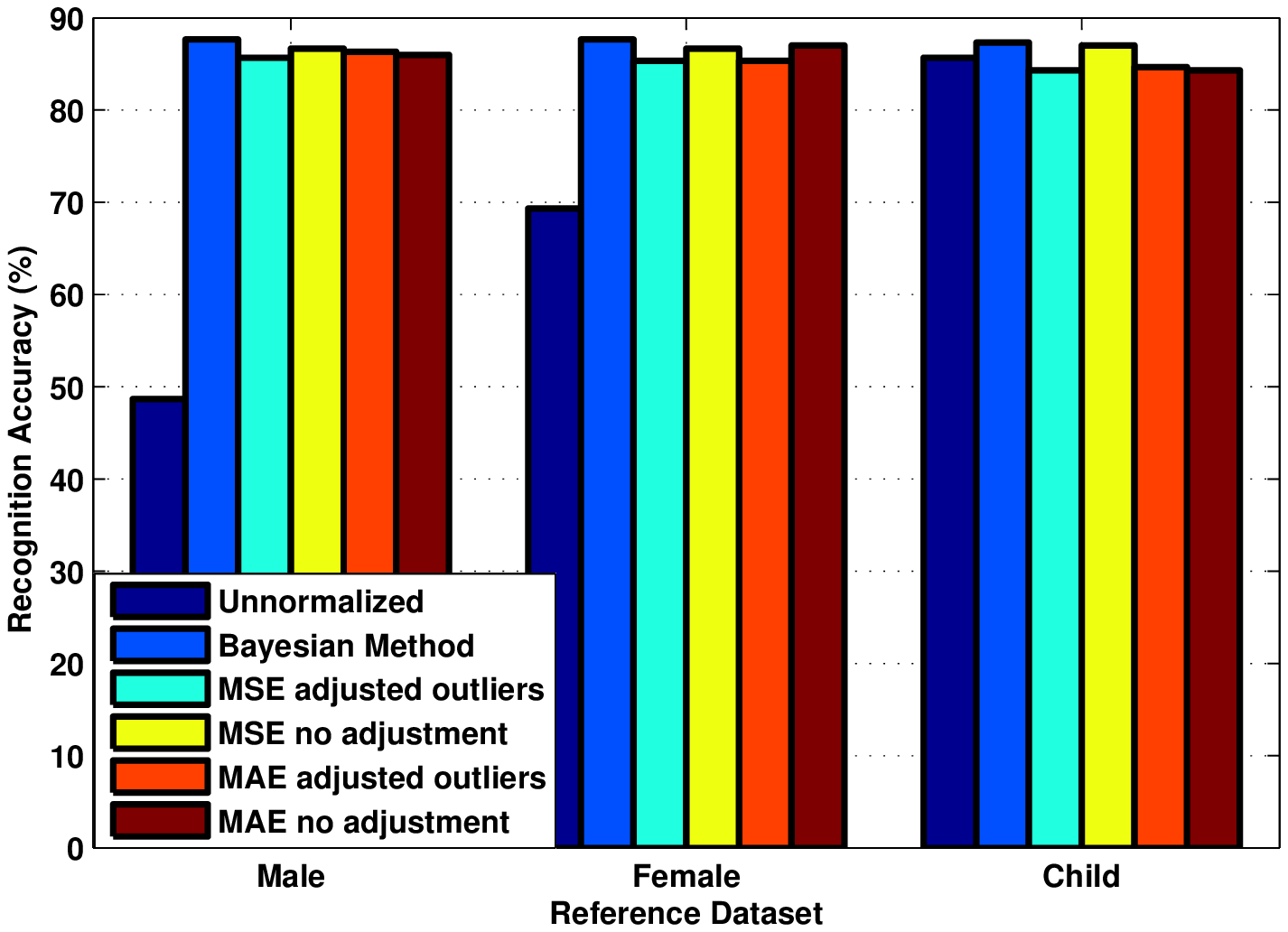}}
  \centerline{(c) Child Subject}
\end{minipage}
\caption{Bar diagrams showing vowel recognition performance on PnB database in terms of Recognition Accuracy for baseline case and using normalization}
\label{pnb_nr_nt}
\end{figure}

\begin{table}[!h]
\begin{center}
 \caption{Table illustrating the Vowel Recognition Performance using normalization in terms of Recognition Accuracy (RA) and its improvement over the baseline case}
 \label{table:vmfcn}
 \begin{adjustbox}{width=\textwidth}
 \begin{tabular}{|c|c|c|c|c|}
 \hline
 \multirow{2}{2.5cm}{\bfseries\stackcell{Normalization\\ Methods}} & \multicolumn{2}{c|}{\bfseries Hillenbrand Database} &
 \multicolumn{2}{c|}{\bfseries Peterson $\&$ Barney Database}\tabularnewline
 \cline{2-5}
 & {\bfseries Recognition Accuracy (\%)} & {\bfseries Improvement(\%)} & {\bfseries Recognition Accuracy (\%)} & {\bfseries Improvement(\%)} \tabularnewline
 \hline
 Baseline Case & 75.2 & - & 83.9 & - \tabularnewline
 \hline
 Bayesian Estimation & 80.1 & 6.5 & 88.6 & 5.6 \tabularnewline
 \hline
 MSE without any adjustment & 75.0  & -0.3  & 82.4  & -1.9 \tabularnewline
 \hline
 MSE with adjusted outliers & 70.2  & -6.6  & 80.1  & -4.5 \tabularnewline
 \hline
 MAE without any adjustment & 78.0 & 3.7  & 84.3  & 0.5 \tabularnewline
 \hline
 MAE with adjusted outliers & 73.3  & -2.5  & 85.6  & 2.0 \tabularnewline
 \hline
 \end{tabular}
 \end{adjustbox}
\end{center}
\end{table}

\subsubsection{Vowel Recognition Performance}

The recognition experiment on vowels can be divided into two categories, namely
\vspace{-1.5mm}
\begin{itemize}
\item {\it Gender Dependent Normalization:}
The recognition accuracies are shown using bar diagrams. Figures \ref{hil_nr_nt} and \ref{pnb_nr_nt} show the vowel recognition performances for Hil and PnB databases respectively. In each Figure, there are three bar diagrams corresponding to different kinds of subject speakers. Each bar diagram comprises of three groups corresponding to different categories of reference speakers. The groups in turn contain 6 bars each, among which the first bar corresponds to baseline case and other bars correspond to various normalization methods.

\item {\it Gender Independent Normalization:}
In this experiment, normalization parameters for a speaker is computed by considering all other speakers present in the database (not of a particular gender) as reference speakers. The recognition accuracy for different experiments are shown in Table \ref{table:vmfcn}. The relative improvements in performance over the baseline case (without normalization) while using normalization, are indicated in a separate column. A negative entry in this column demonstrates a degradation in performance.
\end{itemize}


The experiments discussed in this section signifies that, a greater amount of performance improvement can be achieved for gender dependent normalization compared to gender independent normalization. It also indicates that, normalization parameters estimated using mean absolute error giver better performance compared to mean square error.


\subsection{Experiments on Speech Recognition}
\label{SpRec}
A Hidden Markov Model (HMM) based speech recognizer \cite{rabiner1989tutorial,mariani1989recent} is used here for speech recognition experiments. A brief description is presented on the speech recognizer as well as the database being used and the experimental set-up is discussed. Thereafter the experiments are carried out and recognition performances are shown in terms of word error rate.

\subsubsection{The Recognizer}
The standard filter-bank front end introduced by Davis and Mermelstein \cite{davis1980comparison} (which is conventionally used in HMM based speech recognizer) is modified to incorporate normalization method for feature extraction. The normalization scheme is applied on power spectrum of windowed signal during feature extraction. Normalized features are extracted from this modified front-end signal processor. This normalization process changes the bandwidth as well as the frequency bin values of the spectrum. Due to these reasons, normalized spectrum needs to be modified before feature-extraction.

\begin{itemize}
\item {\it Bandwidth Adjustment:} Depending on the value of $\alpha$, bandwidth of the spectrum changes. For values of $\alpha>1$, bandwidth increases, whereas $\alpha<1$ decreases the bandwidth. This difference in bandwidth is adjusted using piecewise linear warping function. The following function is applied on the warped spectra,
\begin{equation}
G'(f)=
\begin{cases}
    G(f),	    & 0 \leq f \leq f_0 \\	
    \frac{f_{max}-G(f_0)}{f_{max}-f_0}(f-f_0)+G(f_0),	& f_0 \leq f \leq f_{max}
\end{cases}
\end{equation}

where, $f_{max}$ is the maximum frequency present in the signal, $f_0$ is a frequency chosen by the user which falls above the highest significant formant in the speech and $G(f)$ is the warped spectra, which we get by applying our affine model based normalization with parameters estimated using different methods discussed earlier.

\item {\it Frequency Bin Adjustment:} In general, for the spectrum of $G(f)$, the amplitude of the spectrum is available only at specific values of $f$. Due to the bandwidth adjustment discussed earlier, the frequency values could no longer be on those specific values. Also, due to the shift in warping function, some of the frequency points might be missing either in the beginning (for positive shift) or in the end (for negative shift) of the spectrum. A simple linear interpolation method is used here to get amplitudes of the spectrum at those values of $f$ \cite{zhan1997speaker, zhan1997vocal} where amplitude of unnormalized spectrum was defined.
\end{itemize}

The step followed by feature extraction from the acoustic data is training of the HMM models using the features. Subsequently the trained HMM models are used for testing purposes. The training and testing methods are modified as well to incorporate normalization into the recognizer. The process of normalized training and testing is discussed in the following paragraph.
\begin{itemize}

\begin{figure}[b]
 \centering
 \centerline{\includegraphics[width=0.75\linewidth,height=5cm]{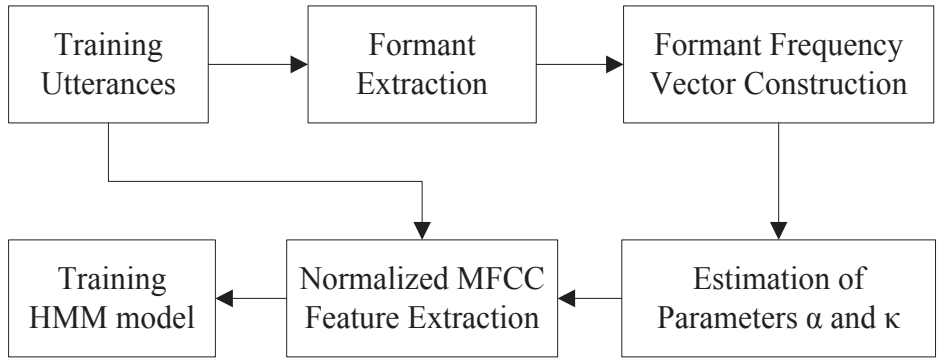}}
 \caption{Block Diagram Illustrating Incorporation of Normalization Method for Training the Recognizer}
\label{fig:train}
\end{figure}

\item {\it Normalized Training:} The training process begins with the computation of formant frequencies from training utterances \cite{snell1993formant}. Then, the formant frequency vector for a speaker is constructed by concatenating formants of all utterances from that speaker. Subsequently, the normalization parameters are estimated using the techniques discussed in Sections \ref{sec:speaker_norm} and \ref{sec:bayesian}. Later normalized MFCC features are extracted from the utterances using normalization parameters which are used to train the HMM model. These steps are summarized in a block diagram shown in Figure \ref{fig:train}. 

\begin{figure}
 \centering
 \centerline{\includegraphics[width=0.75\linewidth,height=6.5cm]{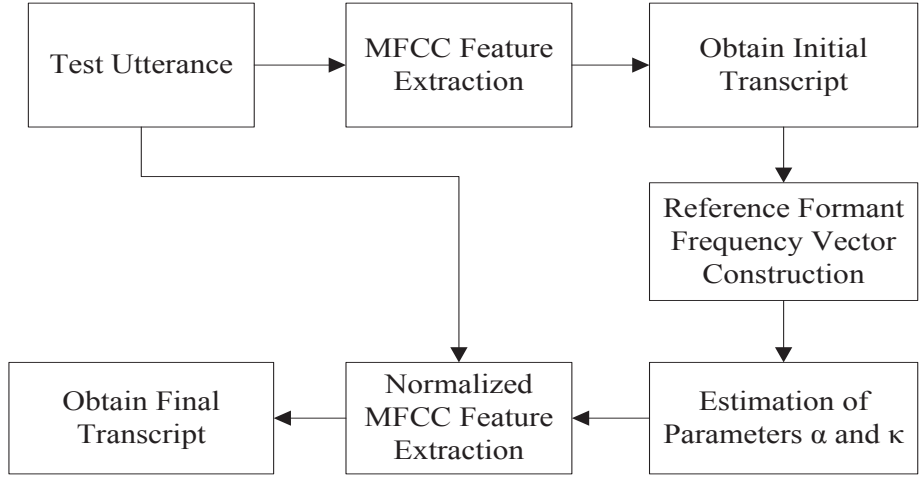}}
 \caption{Block Diagram Illustrating Incorporation of Normalization Method for Testing an Utterance using the Recognizer}
\label{fig:test}
\end{figure}
\item {\it Normalized Testing:} In order to incorporate normalization process into testing utterances a two pass approach through the recognizer is adopted. In the first pass, features are extracted from a test utterance to obtain an initial transcript. Utterances corresponding to initial transcript in the training database helps to construct a class of reference formant frequency vectors for the test utterance under consideration. Subsequently, the normalization parameters are computed using the class of reference formant frequency vectors. These parameters are used to extract normalized MFCC features which are again passed through the recognizer to obtain final transcript. Figure \ref{fig:test} summarizes the steps discussed in this module using block diagram.

\end{itemize}

\subsubsection{Hindi Language Database}

The database used here, is part of a bigger database, collected for a project, funded by Ministry of Communication $\&$ Information Technology, Govt. of India. The goal of the project was to develop a system in which queries regarding prices of some commodities in a particular district can be made through mobile phones using speech modality. A speech database was developed for this purpose, consisting of the names of different commodities and districts. The original database comprises of speech data, collected in six different Indian languages. These languages are spoken in many regional and social dialects, with their own styles. The speech data were collected from about 1000 farmers for each language across different districts to capture the variation in dialects. Farmers were encouraged to do the recordings using their own mobile phones in the environment that they live in. The noise level of data are substantially higher due to the surrounding environment, which can be the field or village country side. The recordings have a sampling rate of 8 kHz and are stored in 16 bit PCM format with a Microsoft .wav header. The challenges posed by this database for the application of speech recognition technology can be summarized as follows,
\begin{itemize}
\item Accent and dialect variations.
\item High level of  background noise.
\item Disfluencies and pauses, since the users are naive.
\item Issues like poor network coverage, interference, fading etc. and their effect on speech data.
\end{itemize}

In this work, the experiments are performed on a subset of the above mentioned database, which are recorded in Hindi language. This subset consists of two groups. In one group, there are 107 commodity names (Commodity Database), and the other group consists of 71 district names (District Database). The Commodity database consists of 7431 utterances [$\sim$38 hrs], whereas District database has 4783 utterances [$\sim$24 hrs].

\begin{table} [!h]
\begin{center}
 \caption{Table of specifications for front end signal processing used in the experiments of speech recognition}
 \begin{tabular}{|c|c|}
 \hline
 {\bf Parameter} & {\bf Default Value} \\ \hline
 Window Length & 20ms \\ \hline
 Filterbank Type & Mel Filterbank \\ \hline
 Number of Mel Filters & 40 \\ \hline
 Number of Cepstra & 13 \\ \hline
 DFT size & 512 \\ \hline
 Lower Filter Frequency & 133.33 Hz \\ \hline
 Upper Filter Frequency & 6855.49 Hz \\ \hline
 Pre-Emphasis Factor & 0.97 \\ \hline
 \end{tabular}
 \label{fesp}
\end{center}
\end{table}

\begin{table}
\begin{center}
 \caption{Table illustrating the Speech Recognition Performance using normalization in terms of Word Error Rate (WER) and its improvement over the baseline case}
 \label{table:tntn}
 \begin{adjustbox}{width=\textwidth}
 \begin{tabular}{|c|c|c|c|c|}
 \hline
 \multirow{2}{2.5cm}{\bfseries\stackcell{Normalization\\ Methods}} & \multicolumn{2}{c|}{\bfseries Commodity Database} &
 \multicolumn{2}{c|}{\bfseries District Database}\tabularnewline
 \cline{2-5}
 & {\bfseries Word Error Rate (\%)} & {\bfseries Improvement(\%)} & {\bfseries Word Error Rate (\%)} & {\bfseries Improvement(\%)} \tabularnewline
 \hline
 Baseline Case & 16.7 & - & 24.6 & - \tabularnewline
 \hline
 Linear VTLN & 15.5 & 7.2 & 22.5 & 8.5 \tabularnewline
 \hline
 Bayesian Estimation & 14.3 & 14.4 & 21.1 & 14.2 \tabularnewline
 \hline
 MSE without any adjustment & 22.4  & -34.1  & 28.3  & -15.1 \tabularnewline
 \hline
 MSE with adjusted outliers & 19.8  & -18.6  & 43.2  & -75.6 \tabularnewline
 \hline
 MAE without any adjustment & 18.6 & -11.4  & 27.5  & -11.8 \tabularnewline
 \hline
 MAE with adjusted outliers & 17.4  & -4.2  & 38.4  & -56.1 \tabularnewline
 \hline
 \end{tabular}
 \end{adjustbox}
\end{center}
\end{table}

\subsubsection{Experimental Conditions}

The databases are divided into two parts, training and testing, as discussed earlier in this section. For Commodity database, the training data consists of 5899 utterances [$\sim$30.5 hrs] and testing data consists of 1532 utterances [$\sim$7.5 hrs], whereas the training and testing databases for District database consists of 3798 [$\sim$20 hrs] and 985 [$\sim$5 hrs] utterances respectively.  The specifications for front end signal processing for feature extraction is given in Table \ref{fesp}. Experiments using context independent phonemes have been carried out. The phonemes in each database are represented using three state left to right HMM. Further, each state of the HMM is modelled using a mixture of 16 Gaussian densities. The experiments are performed using the Sphinx3 toolkit. Note that the conducted experiments are gender independent i.e. the normalization parameters are extracted without gender information of the speaker. This method was adopted to increase usability of the system in real life scenario where such information may not be available.

\subsubsection{Speech Recognition Performance}
The recognition performance for different experiments is evaluated using Word Error Rate (WER). The results of these experiments are shown in Table \ref{table:tntn}. First row of this table corresponds to the baseline case, when there is no speaker normalization being used. The following row shows recognition performance using linear normalization model \cite{lee1996speaker}. All the following rows display performance using affine model. The parameters of this affine model is estimated using various techniques discussed in Sections \ref{sec:speaker_norm} and \ref{sec:bayesian}. Performance using normalization is compared with baseline case and relative improvements are indicated in a separate column. A negative entry in this column signifies performance degradation. Note that, background noise can significantly degrade the recognition performance. This fact was presented in a work by Hirsch and Pearce, which can be found in \cite{hirsch2000aurora}. In the standard Aurora-2 task for recognising the ten digits and `oh' in American English (i.e. only 11 word vocabulary), they have shown that the performance can vary from 99\% to 10\% depending on the background noise. In this work also, the speech data in both databases contain large amount of background noise which justifies high WER. The results using normalization indicate that, changing the normalization model from linear to affine, performance can be improved. The parameter estimation method plays an important role in determining the recognition performance using the affine model, and the effect can be easily observed in recognition results presented in Table \ref{table:tntn}. Among all estimation techniques used, only Bayesian estimation method improves performance of recognizer. This improvement is almost twice than that of using linear model.

%% file: conclusion.tex
\section{Conclusion}
\label{sec:Con}
A Bayesian approach to speaker normalization is proposed in this paper. The variation of vocal tract length among different speakers is modelled using an affine model. The parameters of the model are estimated using Error Function Minimization technique and its limitations are discussed. Subsequently, a Bayesian method of parameter estimation is proposed and the framework is described for the problem under consideration. A special type of Markov Chain Monte Carlo method called Gibbs Sampler is used to implement the Bayesian estimation. The need for hyperparameter estimation is also presented. Later, maximum likelihood estimation is used to estimate the hyperparameters corresponding to model parameters.

A Mahalanobis distance based vowel recognizer is introduced and used for the experiments on vowel normalization. First three formant frequencies of a vowel are considered in this kind of recognizer. Experiments are performed for both Gender dependent as well as Gender independent normalization. It is observed that, the improvement in performance in case of gender dependent normalization is higher than that of gender independent normalization. This indicates that, the proposed approach is better suited for cross-gender normalization. The normalization scheme is further used for speech recognition experiments using Hidden Markov Models. Techniques like bandwidth and frequency bin adjustment is used for this purpose. The proposed normalization method requires prior knowledge about the transcript of the utterance under test. Therefore, a two pass approach through the recognizer is proposed to solve this problem. All the experiments discussed in this paper justifies that, Bayesian estimation method gives better performance compared to other methods.

The prior distributions used for all experiments in this work are non-informative, data-driven priors. Currently, methods that use non-Gaussian priors and informative priors are being explored to estimate the speaker normalization parameters. Additionally, the possibility of using higher order speaker normalization models in the proposed Bayesian parameter estimation framework is also being investigated.

%% file: appendix.tex
\appendix
\subsection{Derivation of Posterior Distribution of $\kappa$ given in Equation (17):}
\label{app:a}

We have $ f_{y}({\bf Y}_i|\kappa, \alpha_{i}, \sigma) $ given in Equation (\ref{eq:priorK}) and the prior of $\kappa$ is Gaussian i.e. $ f_1(\kappa | a, b) = \frac{1}{\sqrt{2\pi}b} e^ \frac{-(\kappa-a)^2}{b^2} $. Let ${\bf Y} = ({\bf Y}_1, {\bf Y}_2 \ldots, {\bf Y}_n)$ and ${\boldsymbol \alpha} = (\alpha_{1}, \alpha_{2}, \ldots, \alpha_{n})$. The joint distribution of $ ({\bf Y}, \kappa | {\boldsymbol \alpha}, \sigma) $ is given by,

\begin{equation} \label{app1}
 \begin{aligned}
 f_{y}^{(1)}({\bf Y}, \kappa | {\boldsymbol \alpha}, \sigma) & = \prod_{i=1}^{n} f_{y}({\bf Y}_i|\kappa, \alpha_{i}, \sigma) f_1(\kappa) \\
 & = \frac{1}{\displaystyle (2 \pi \sigma^2)^{nr/2}} e^{\frac{-\sum_{i=1}^n ({\bf Y}_i-{\bf \mu}_i)^T({\bf Y}_i-{\bf \mu}_i)}  {2\sigma^2}} \frac{1}{\sqrt{2\pi}b} e^ \frac{-(\kappa-a)^2}{b^2} \\
 & = \frac{1}{(2\pi)^{\frac{nr+1}{2}} b \sigma^{nr}} e^{-(\kappa^2 A_{\kappa} -2\kappa B_{\kappa} + C_{\kappa})} \\
 & = \frac{1}{(2\pi)^{\frac{nr+1}{2}} b \sigma^{nr}} e^{-A_{\kappa}(\kappa - \frac{B_{\kappa}}{A_{\kappa}})^2 +  (\frac{B_{\kappa}^2}{A_{\kappa}} - C_{\kappa})}
 \end{aligned}
\end{equation}

where,
$ A_{\kappa} = \frac{1}{\frac{r}{2\sigma^2} \sum_{i=1}^n (\alpha_i-1)^2 + \frac{1}{2b^2}} $
\vspace{2mm}

\hspace{10.5mm}$ B_{\kappa} = \frac{a}{2b^2} + \frac{1}{2\sigma^2}\sum_{i=1}^n (\alpha_i-1) \{({\bf Y}_i-\alpha_i{\bf X})^T. {\bf 1}\} $
\vspace{2mm}

\hspace{10.5mm}$ C_{\kappa} = \frac{a^2}{2b^2} + \frac{1}{2\sigma^2}\sum_{i=1}^n \{({\bf Y}_i-\alpha_i{\bf X})^T ({\bf Y}_i-\alpha_i{\bf X})\} $
\vspace{2mm}

Now the joint distribution of $ ({\bf Y} | \alpha, \sigma) $ can be obtained by integrating  Equation (\ref{app1}) over $\kappa$ as follows,

\begin{equation} \label{app2}
\begin{aligned}
f_{y}^{(2)}({\bf Y} | {\boldsymbol \alpha}, \sigma) & = \int_{-\infty}^{\infty}f_{y}^{(1)}({\bf Y}, \kappa | {\boldsymbol \alpha}, \sigma)\, d\kappa\ \\
& = \frac{1}{(2\pi)^{\frac{nr+1}{2}} b \sigma^{nr}} \int_{-\infty}^{\infty} e^{-A_{\kappa}(\kappa - \frac{B_{\kappa}}{A_{\kappa}} )^2 + (\frac{B_{\kappa}^2}{A_{\kappa}} - C_{\kappa})}\, d\kappa\ \\
& = \frac{1}{(2\pi)^{\frac{nr+1}{2}} b \sigma^{nr}} \left(\sqrt{\frac{\pi}{A_{\kappa}}}\right) e^{(\frac{B_{\kappa}^2}{A_{\kappa}} - C_{\kappa})}
\end{aligned}
\end{equation}

Finally, posterior distribution of $\kappa$ can be obtained by dividing Equation (\ref{app1}) by Equation (\ref{app2})
\begin{equation} \label{app3}
\begin{aligned}
f_{\kappa}(\kappa | {\bf Y}, {\boldsymbol \alpha}, \sigma) & = \frac{f_{y}^{(1)}({\bf Y}, \kappa | {\boldsymbol \alpha}, \sigma)}{f_{y}^{(2)}({\bf Y} | {\boldsymbol \alpha}, \sigma)} \\
& = \left(\sqrt{\frac{A_{\kappa}}{\pi}}\right) e^{-A_{\kappa}(\kappa - \frac{B_{\kappa}}{A_{\kappa}})^2} \\
& = \frac{1}{\sqrt{2\pi\sigma_{post}^2}} e^{\frac{-(\kappa-\mu_{post})^2}{2\sigma_{post}^2}}
\end{aligned}
\end{equation}

where, $ \sigma^2_{\kappa} = \frac{1}{2A_{\kappa}}$ and $\mu_{\kappa} = \frac{B_{\kappa}}{A_{\kappa}} $ \\

\subsection{Derivation of Posterior Distribution of $\alpha_{i}$ given in Equation (18):}
\label{app:b}

The prior distribution of $\alpha_i$ is assumed to be Gaussian i.e. $ f_1(\alpha_i | c, d) = \frac{1}{\sqrt{2\pi}d} e^ \frac{-(\alpha_i-c)^2}{d^2} $. So, the joint distribution of ${\bf Y}_i$ and $\alpha_{i}$ can be obtained by multiplying $f_1(\alpha_i)$ with $f_{y}({\bf Y}_i|\kappa, \alpha_{i}, \sigma)$ as follows,

\begin{equation} \label{app4}
\begin{aligned}
f_{\alpha}^{(1)}({\bf Y}_i, \alpha_{i}|\kappa, \sigma) & = f_{y}({\bf Y}_i|\kappa, \alpha_{i}, \sigma) f_1(\alpha_i) \\
& = \frac{1}{\displaystyle (2 \pi \sigma^2)^{r/2}} e^{\frac{-({\bf Y}_i-{\bf \mu}_i)^T({\bf Y}_i-{\bf \mu}_i)}{2\sigma^2}} \frac{1}{\sqrt{2\pi}d} e^ \frac{-(\alpha_i-c)^2}{d^2} \\
& = \frac{1}{(2\pi)^{\frac{r+1}{2}} d \sigma^{r}} e^{-(A_{\alpha_i} \alpha_i^2 -2B_{\alpha_i} \alpha_i + C_{\alpha_i})} \\
& = \frac{1}{(2\pi)^{\frac{r+1}{2}} d \sigma^{r}} e^{-A_{\alpha_i}(\alpha_i - \frac{B_{\alpha_i}}{A_{\alpha_i}} )^2 + (\frac{B_{\alpha_i}^2}{A_{\alpha_i}} - C_{\alpha_i})}
\end{aligned}
\end{equation}

where, $ A_{\alpha_i} = \frac{1}{2d^2} + \frac{{\bf X}^T {\bf X} + 2 \kappa ({\bf X}^T {\bf 1}) + r \kappa^2}{2\sigma^2} $
\vspace{2mm}

\hspace{10.5mm}$ B_{\alpha_i} = \frac{{\bf X}^T {\bf Y}_i + \kappa ({\bf X} + {\bf Y}_i)^T {\bf 1} + r \kappa^2}{2\sigma^2} + \frac{c}{2d^2} $
\vspace{2mm}

\hspace{10.5mm}$ C_{\alpha_i} = \frac{{\bf Y}_i^T {\bf Y}_i + 2\kappa ({\bf Y}_i^T {\bf 1}) + r \kappa^2}{2\sigma^2} + \frac{c^2}{2d^2} $
\vspace{2mm}

Now the distribution of $ ({\bf Y}_i |\kappa, \sigma) $ can be obtained by integrating  Equation (\ref{app4}) over $\alpha_i$ as follows,

\begin{equation} \label{app5}
\begin{aligned}
f_{\alpha}^{(2)}({\bf Y}_i |\kappa, \sigma) & = \int_{-\infty}^{\infty} f_{\alpha}^{(1)}({\bf Y}_i, \alpha_{i}|\kappa, \sigma)\, d\alpha_i\ \\
& = \frac{1}{(2\pi)^{\frac{r+1}{2}} d \sigma^{r}} \int_{-\infty}^{\infty} e^{-A_{\alpha_i}(\alpha_i - \frac{B_{\alpha_i}}{A_{\alpha_i}})^2 + ( \frac{B_{\alpha_i}^2}{A_{\alpha_i}} - C_{\alpha_i})}\, d\alpha_i\ \\
& = \frac{1}{(2\pi)^{\frac{r+1}{2}} d \sigma^{r}} \left(\sqrt{\frac{\pi}{A_{\alpha_i}}}\right) e^{( \frac{B_{\alpha_i}^2}{A_{\alpha_i}} - C_{\alpha_i})}
\end{aligned}
\end{equation}

Finally, posterior distribution of $\alpha_{i}$ can be obtained by dividing Equation (\ref{app4}) by Equation (\ref{app5})

\begin{equation} \label{app6}
\begin{aligned}
f_{\alpha}(\alpha_{i}|\kappa,\sigma,{\bf Y}_i) & = \frac{f_{\alpha}^{(1)}({\bf Y}_i, \alpha_{i}|\kappa, \sigma)}{f_{\alpha}^{(2)}({\bf Y}_i |\kappa, \sigma)} \\
& = \left(\sqrt{\frac{A_{\alpha_i}}{\pi}}\right) e^{-A_{\alpha_i}(\kappa - \frac{B_{\alpha_i}}{A_{\alpha_i}})^2} \\
& = \frac{1}{\sqrt{2\pi\sigma_{post}^2}} e^{\frac{-(\alpha_i-\mu_{post})^2}{2\sigma_{post}^2}}
\end{aligned}
\end{equation}

where, $ \sigma^2_{\alpha_i} = \frac{1}{2A_{\alpha_i}}$ and $\mu_{\alpha_i} = \frac{B_{\alpha_i}}{A_{\alpha_i}} $

\subsection{Derivation of Posterior Distribution of $\sigma$ given in Equation (19):}
\label{app:c}

The prior distribution of $\sigma$ is assumed to be uniform i.e. $ f_2(\sigma | \theta_1, \theta_2) = \frac{1}{\theta_2 - \theta_1} $. So, the joint distribution of ${\bf Y}_i$ and $\sigma$ can be obtained by multiplying $f_2(\sigma | \theta_1, \theta_2)$ with $f_{y}({\bf Y}_i|\kappa, \alpha_{i}, \sigma)$ as follows,

\begin{equation} \label{app7}
\begin{aligned}
f_{\sigma}^{(1)}({\bf Y}, \sigma | {\boldsymbol \alpha}, \kappa) & = \prod_{i=1}^{n} f_{y}({\bf Y}_i|\kappa, \alpha_{i}, \sigma) f_2(\sigma) \\
& = \frac{1}{\displaystyle (2 \pi \sigma^2)^{nr/2}} e^{\frac{-\sum_{i=1}^n ({\bf Y}_i-{\bf \mu}_i)^T({\bf Y}_i-{\bf \mu}_i)}{2\sigma^2}} \frac{1}{\theta_2 - \theta_1} \\
& = \frac{1}{(2\pi)^{\frac{nr}{2}} (\theta_2 - \theta_1) \sigma^{nr}} e^{\frac{-\sum_{i=1}^n ({\bf Y}_i-{\bf \mu}_i)^T({\bf Y}_i-{\bf \mu}_i)}{2\sigma^2}}
\end{aligned}
\end{equation}

Now the distribution of $ ({\bf Y} | {\boldsymbol \alpha}, \kappa) $ can be obtained by integrating  Equation (\ref{app7}) over $\sigma$ as follows,

\begin{equation} \label{app8}
\begin{aligned}
f_{\sigma}^{(2)}({\bf Y} | {\boldsymbol \alpha}, \kappa) & = \int_{-\infty}^{\infty} f_{\sigma}^{(1)}({\bf Y}, \sigma | {\boldsymbol \alpha}, \kappa)\, d\sigma \\
& = \int_{-\infty}^{\infty} \frac{1}{(2\pi)^{\frac{nr}{2}} (\theta_2 - \theta_1) \sigma^{nr}} e^{\frac{-\sum_{i=1}^n ({\bf Y}_i-{\bf \mu}_i)^T({\bf Y}_i-{\bf \mu}_i)}{2\sigma^2}}\, d\sigma
\end{aligned}
\end{equation}

Finally, posterior distribution of $\alpha_{i}$ can be obtained by dividing Equation (\ref{app7}) by Equation (\ref{app8})

\begin{equation} \label{app9}
\begin{aligned}
f_{\sigma}(\sigma | \kappa, {\boldsymbol \alpha}, {\bf Y}) & = \frac{f_{\sigma}^{(1)}({\bf Y}, \sigma | {\boldsymbol \alpha}, \kappa)}{f_{\sigma}^{(2)}({\bf Y} | {\boldsymbol \alpha}, \kappa)} \\
& = \frac{\frac{e^{-\frac{1}{2\sigma^2}\sum_{i=1}^n({\bf Y}_i-{\bf \mu}_i)^T({\bf Y}_i-{\bf \mu}_i)}}{\sigma^{nr}}} {\int_{\theta_1}^{\theta_2} \frac{e^{-\frac{1}{2\sigma^2}\sum_{i=1}^n({\bf Y}_i-{\bf \mu}_i)^T({\bf Y}_i-{\bf \mu}_i)}  }{\sigma^{nr}} d\sigma}
\end{aligned}
\end{equation}

Assuming $\beta = \frac{1}{2}\sum_{i=1}^n({\bf Y}_i-{\bf \mu}_i)^T({\bf Y}_i-{\bf \mu}_i)$ the denominator in the above Equation can be written as,

\begin{equation} \label{app10}
\begin{aligned}
{\int_{\theta_1}^{\theta_2} \frac{e^{-\frac{\beta}{\sigma^2}}}{\sigma^{nr}} d\sigma} & = \frac{1}{2} \beta^{\frac{1-nr}{2}} \int_\frac{\beta}{\theta_1^2}^\frac{\beta}{\theta_1^2} {z^{\frac{nr-1}{2} - 1} e^{-z}}\, dz \\
& = \frac{1}{2} \beta^{\frac{1-nr}{2}} \left(\gamma-\gamma_l-\gamma_u\right)
\end{aligned}
\end{equation}

where, $ \gamma = \Gamma\left(\frac{nr-1}{2}\right) $, $ \gamma_l = \Gamma_{lower}\left(\frac{\beta}{{\theta_2}^2},\frac{nr-1}{2}\right) \hspace{5mm} and \hspace{5mm} \gamma_u = \Gamma_{upper}\left(\frac{\beta}{{\theta_1}^2},\frac{nr-1}{2}\right)$

\subsection{Derivation of Integrated Likelihood Function given in Equation (28):}
\label{app:d}

The likelihood function of $\Theta$ given in Equation (\ref{eq:likeli}) can be written as follows,

\begin{equation} \label{app11}
\begin{aligned}
 L(\Theta|\Omega,{\bf Y}) & = f_{1}(\kappa|\Theta_{\kappa})f_{2}(\sigma|\Theta_{\sigma})\prod_{i=1}^{n}f_{y}({\bf Y}_i|\kappa,\alpha_i,\sigma)f_{1}(\alpha_i|\Theta_{\alpha}) \\
 & = \left(\frac{1}{\sqrt{2\pi}b} e^ \frac{-(\kappa-a)^2}{b^2} \frac{1}{\theta_2 - \theta_1}\right)
  \left(\prod_{i=1}^{n} \frac{1}{\displaystyle (2 \pi \sigma^2)^{r/2}} e^{\frac{-({\bf Y}_i-{\bf \mu}_i)^T({\bf Y}_i-{\bf \mu}_i)}{2\sigma^2}} \frac{1}{\sqrt{2\pi}d} e^ \frac{-(\alpha_i-c)^2}{d^2}\right)
\end{aligned}
\end{equation}

Equation (\ref{app11}) is integrated over ${\boldsymbol \alpha}$, $\kappa$ and $\sigma$ to obtain an integrated likelihood function solely dependent on hyperparameters as given below,

\begin{equation} \label{app12}
\begin{aligned}
IL(\Theta) & = \int_{\theta_1}^{\theta_2} \frac{1}{\theta_2 - \theta_1} \left( \int_{-\infty}^{\infty} \frac{1}{\sqrt{2\pi}b} e^ \frac{-(\kappa-a)^2}{b^2} \left( \prod_{i=1}^{n} \int_{-\infty}^{\infty} \frac{e^{ \frac{-({\bf Y}_i-{\bf \mu}_i)^T({\bf Y}_i-{\bf \mu}_i)}{2 \sigma^2} + \frac{-(\alpha_i-c)^2}{2 d^2} }} {(2 \pi \sigma^2)^{r/2} \sqrt{2\pi} d}\,  d \alpha_i  \right)\, d\kappa \right)\, d\sigma \\
& = \int_{\theta_1}^{\theta_2} \frac{1}{\theta_2 - \theta_1} \left(\int_{-\infty}^{\infty} \frac{f(\kappa,\:\sigma,\:c,\:d)}{\sqrt{2\pi}b} e^{ \frac{-(\kappa-a)^2}{2 b^2} }  d\kappa \right) d\sigma
\end{aligned}
\end{equation}

\vspace{3mm}
The function, $f(\kappa,\:\sigma,\:c,\:d)$ in Equation (\ref{app12}) is calculated as follows,
\begin{equation} \label{app13}
\begin{aligned}
f(\kappa,\:\sigma,\:c,\:d) & = \prod_{i=1}^{n} \int_{-\infty}^{\infty} \frac{e^{ \frac{-({\bf Y}_i-{\bf \mu}_i)^T({\bf Y}_i-{\bf \mu}_i)}{2 \sigma^2} + \frac{-(\alpha_i-c)^2}{2 d^2} }} {(2 \pi \sigma^2)^{r/2} \sqrt{2\pi} d}\,  d \alpha_i \\
& = \prod_{i=1}^{n} \frac{1}{(2\pi)^{\frac{r+1}{2}} d \sigma^{r}} \left(\sqrt{\frac{\pi}{A_{\alpha_i}}}\right) e^{( \frac{B_{\alpha_i}^2}{A_{\alpha_i}} - C_{\alpha_i})} \\
& = \frac{A_{\alpha_i}^{\frac{-n}{2}}}{2^{\frac{n(r+1)}{2}}\pi^{\frac{nr}{2}} d^n \sigma^{nr}} e^{\sum_{i=1}^{n}\left( \frac{B_{\alpha_i}^2}{A_{\alpha_i}} - C_{\alpha_i}\right)}
\end{aligned}
\end{equation}